\documentclass[preprints,article,accept,pdftex,moreauthors]{Definitions/mdpi}

\usepackage{amsmath}
\usepackage{amsthm}
\usepackage{amssymb}
\usepackage{amsfonts}
\usepackage{bm}
\usepackage[overload]{empheq} 
\usepackage{fix-cm} 
\usepackage{physics}
\usepackage{tikz}
\usepackage{tikz-cd}
\usepackage{verbatim}

\firstpage{1} 
\pubvolume{1}
\issuenum{1}
\articlenumber{0}
\pubyear{2024}
\copyrightyear{2024}
\datereceived{ } 
\daterevised{ } 
\dateaccepted{ } 
\datepublished{ } 
\hreflink{https://doi.org/} 

\Title{A perturbative approach to the solution of the Thirring quantum cellular automaton}

\TitleCitation{A perturbative approach to the solution of the Thirring quantum cellular automaton}

\Author{Alessandro Bisio $^{1}$\orcidA{}, Paolo Perinotti $^{2}$\orcidB{}, Andrea Pizzamiglio $^{3}$ and Saverio Rota $^{4}$}

\AuthorNames{Alessandro Bisio, Paolo Perinotti, Andrea Pizzamiglio and Saverio Rota}

\AuthorCitation{Bisio, A.; Perinotti, P.; Pizzamiglio, A.; Rota, S.}

\address{%
$^{1}$ \quad Dipartimento di Fisica, Università di Pavia and INFN sezione di Pavia; alessandro.bisio@unipv.it\\
$^{2}$ \quad Dipartimento di Fisica, Università di Pavia and INFN sezione di Pavia; paolo.perinotti@unipv.it\\
$^{3}$ \quad Dipartimento di Fisica, Università di Pavia and INFN sezione di Pavia; andrea.pizzamiglio01@universitadipavia.it\\
$^{4}$ \quad Dipartimento di Fisica, Università di Pavia; saverio.rota01@universitadipavia.it
}

\abstract{The Thirring Quantum Cellular Automaton (QCA) describes the discrete time dynamics of local fermionic modes that evolve according to one step of the Dirac cellular automaton followed by the most general on-site number-preserving interaction, and serves as the QCA counterpart of the Thirring model in quantum field theory. 
In this work, we develop perturbative techniques for the QCA path-sum approach, expanding both the number of interaction vertices and the mass parameter of the Thirring QCA. By classifying paths within the regimes of very light and very heavy particles, we computed the transition matrices in the two- and three-particle sectors to the first few orders. Our investigation into the properties of the Thirring QCA, addressing the combinatorial complexity of the problem, yielded some useful results applicable to the many-particle sector of any on-site number-preserving interactions in one spatial dimension.}


\begin{document}


\section{Introduction}\label{introduction}
Understanding the dynamics of interacting many-body quantum systems remains one of the paramount unsolved challenges in physics. The intricate behaviours and interdependencies within these systems render both analytical descriptions and numerical simulations exceedingly arduous.  Nonetheless, the past decades of research have heralded an uplifting new horizon. The confluence of cutting-edge quantum technologies and the maturation of quantum information theory offers a wealth of promising investigative tools, both practical and theoretical. The rapidly increasing precision in control over quantum apparatus suggests that we will soon be able to test both novel and long-standing hypotheses, ranging from simulations in particle physics and condensed matter to exploring the very foundations of contemporary physics.

Specifically, in addressing the inefficiency of classical computers in simulating quantum systems, Quantum Cellular Automata (QCAs) delineate practical quantum algorithms for simulating the dynamics of an interacting quantum many-body system \cite{feynman1981simulating,deutsch1985quantum,qcagrozei,Farrelly_2020,preskill2018quantum}. Quantum cellular automata are homogeneous networks of local quantum systems that abide by a translationally invariant discrete-time evolution, each interacting with a finite number of neighbours. The most interesting QCAs for quantum simulations are those that can be implemented as Finite-Depth Quantum Circuits (FDQC). Furthermore, QCAs offer an inherent architecture for the implementation of quantum simulation hardware and for distributed quantum computation \cite{arrighi2019overview,Farrelly_2020}, and they are universal for quantum computation as they were proved equivalent both to the quantum Turing machine and the circuit paradigm of quantum computation \cite{arrighi2019overview}. 

In recent years, QCAs, and Quantum Walks (QWs)---which can be regarded as the one-particle sector, or first quantisation counterpart of QCAs---have been investigated as instruments for simulating relativistic quantum fields and as discrete methods for studying the foundations of Quantum Field Theory (QFT) \cite{PhysRevD.49.6920,beny2017inferring,osborne2019continuum,Perinotti_2020,mlodinow2021fermionic,zimboras2022does,eon2023relativistic}. There are many advantages of such an approach to the foundations. From an information theoretical standpoint, quantum systems might be conceived as elementary information carriers, rather than elementary constituents of matter, circumventing the molasses of interpretational issues of quantum mechanics. With this premises, relationships between elementary quantum systems can be thought of as  logical connections within an algorithm, as opposed to the usual space-time relations. Along this line, embracing the principle initially proposed by Feynman \cite{feynman1981simulating} and later refined by Deutsch \cite{deutsch1985quantum}, that every experiment involving a finite space-time region should be perfectly simulated by a finite quantum algorithm, we can devise physical laws as algorithms that govern the update of memory registers constituting a physical system. Within this scope, the QCA framework naturally emerges by imposing---once reframed in this context---the characteristic properties of physical laws onto such algorithms, as locality and homogeneity \cite{PhysRevA.90.062106}.

This approach to fundamental physics has relevant practical advantages. Over the last decades, several quantum simulation schemes have been devised \cite{georgescu2014quantum}, yet these are predominantly Hamiltonian-based. They rely on a discrete-space continuous-time version of the QFT. Analog quantum systems mimicking the Hamiltonian are sought, or alternatively, staggered trotterisation (Suzuki-Trotter decomposition) is performed to obtain unitary transformations that may be implemented as a quantum circuit on a digital quantum computer. Trotterised gates are close to the identity though and contrast with QCAs gates, which conversely, not being subject to trotterisation conditions, can be highly non-trivial, converging faster to physical dynamics \cite{Arrighi_2020, farrelly2020discretizing,bisio2021scattering} under suitable conditions. Furthermore, the possibility of framing algorithms not originating from conventional QFTs suggests a broader spectrum of possibilities for simulating them via quantum computers beyond lattice QFT. 

In the second place, given that the traditional action-based approach is only surreptitiously continuous, namely up to renormalisation \cite{wilson1983renormalization}, a crucial procedure for both its mathematical amenability and physical significance, it is conceptually clearer to start from new premises, based on a natively discrete theory.  Indeed, QCAs exhibit manifest unitarity and possess a local formulation. 

Gauge invariance can be handled as outlined in \cite{Arrighi_2018,arnault2016quantum,centofanti2024}. Moreover, an ab initio quantum description, namely not obtained by carrying out a quantization procedure on a classical theory, sidesteps the ambiguities that arise when striving to develop a quantum theory based on a classical action, and due to the emergent nature of the geometry of space-time, it represents an enticing arena for approaching a quantum theory of gravity \cite{bibeau2015doubly,Bisio_2016,arrighi2017quantum,apadula2020symmetries}.

There are of course also drawbacks in facing a new formulation of QFT. In the first place, 
being intrinsically discrete, this kind of models were for a long time dismissed as 
incompatible with the relativistic framework that is the crucial feature of quantum 
fields~\cite{PhysRev.73.414}. However, in recent years their compatibility with Lorentz-covariance has been explicitly demonstrated in compliance with different prescriptions  \cite{Bisio_2016,debbasch2019action,arrighi2014discrete}, e.g. by identifying a change of inertial frame with a symmetry of the dynamics. The bounded speed of light in circuit-based quantum simulation is naturally enforced by the wiring between local quantum gates. 

Secondarily, despite QCAs have achieved promising success in recovering \emph{free} quantum field theories from principles of information processing \cite{Perinotti_2020}, to date, a thorough study of interaction processes within the QCA framework is lacking, representing a largely unexplored avenue. Like in QFT, this point is the source of main troubles, and the newborn theory of QCAs cannot yet count on the wealth of techniques and 
approaches of the standard formulation. This requires an extensive review of theoretical tools with the aim of either adapting them to the discrete framework or to replace them by new, more suitable techniques.

In the present study, we tackle the issue by endeavouring to expand a specific instance of the QCA approach that has proved successful in solving the non-interacting dynamics: the \emph{path sum} \cite{dariano2014pathint}. As a case study, we consider the (1+1)-dimensional Thirring Quantum Cellular Automaton. The interaction described by the Thirring QCA is the most general one, being both number-preserving---meaning the number of particles remains constant before and after the interaction---and on-site, requiring particles to occupy the same lattice site simultaneously to interact. Moreover, our choice is driven by three primary reasons. In the first place, the interaction-free model has been exactly treated via a path-sum approach~\cite{dariano2014pathint}, providing a solid starting toolbox, and the interacting one has been analytically solved in the two-particle sector~\cite{bisio2018thirring}, allowing us to validate the results and check the accuracy of approximations. Secondarily, in the (1+1)-dimensional case, we can easily visualise the discrete evolution of the automaton on a two-dimensional space-time lattice. The evolution of a particle with specified initial and final conditions can be effectively represented by a collection of interfering paths connecting these points. Within the path-sum paradigm, transition amplitudes are then obtained by summing over such paths.  
Finally, the Thirring model in QFT belongs to the restricted category of integrable models. Though it is not completely obvious that the Thirring QCA obtains the Thirring model in some limit, this allows us to borrow techniques from established results, and to have a source for future validation and developments. 

Despite the simplicity of local rules governing QCAs, the emerging behaviour exhibits increasing complexity as the evolution progresses. While this is one of the general reasons why QCAs represent a valuable tool in studying complex systems, it is also the reason why their actual classification proves to be remarkably challenging~\cite{gross2012index,PhysRevB.99.085115,freedman2020classification,haah2022nontrivial}. Within the path sum approach, and given the on-site nature of Thirring interaction, the origin of the complexity here is embodied in a comprehensive classification of all possible paths based on the number of intersections that yield an interaction. Such classification requires to take into account a both the Pauli exclusion principle and the fact that the free dynamics of the QCA univocally determines the internal state of the systems at every lattice site. 

At the outset, one is led to follow a perturbative approach in the number of interactions, where fixed the number of time steps, the initial and final conditions, and we expanded transition amplitudes in a sum of terms, each corresponding to a fixed the number of interactions. Unfortunately, the algorithm that identifies all paths with a given number of interactions (i.e.~crossings) is computationally expensive. Consequently, we focused on a different perturbative approach, consisting in a power expansion in the mass parameter of the automaton. The regularity of the paths within the extremal regimes of ultra-light and ultra-heavy fermions, allowed for their full enumeration. We then computed the transition matrices in these regimes up to third order and up to the three particles sector. 

The paper is structured as follows.
Section \ref{diracqca} gives an outline of the one-dimensional Dirac QCA \cite{BISIO2015244}. We then provide an overview of the path sum approach to its solution \cite{dariano2014pathint}. 
In section \ref{thirring} a synopsis of the Thirring QCA \cite{bisio2018thirring} is presented. 
Section \ref{mathtools} shows an overarching characterization of the particle paths, grounded in the general features of Thirring QCA. Sections \ref{pert_int} and \ref{pert_mass} cover the path sum approach in the two-particle sector, with a perturbative technique in the number of interactions and in the particle mass respectively. Finally, in section \ref{three} we discuss the multi-particle sector and in particular the three-particle case.
\section{One-dimensional Dirac QCA}
\label{diracqca}

\subsection{Derivation}
The following discussion may appear as a mere exercise, yet it unveils the potential to reformulate the description of a relativistic quantum theory without initially resorting to the conventional categories of physics, such as spacetime or differential equations thereof. This proves that is feasible to develop an emergent physical semantics on top of the concept of information and its processing, without introducing theoretical physical concepts ab initio. The operational physical notions will be connected to the mathematical symbols through comparison with observations, in a similar way as thermodynamic quantities are derived in statistical mechanics.

Indeed, the Dirac equation can be derived solely from fundamental principles of information processing, without appealing a priori to a spacetime background or special relativity. It has been shown that the Dirac equation in any spatial dimension can be recovered from the large-scale dynamics of a QCA of fermionic systems satisfying linearity, unitarity, locality, homogeneity, and discrete isotropy. Here we briefly sketch the construction (see \cite{PhysRevA.90.062106} for a full derivation). The systems are local fermionic modes labelled by the elements $g$ of a countable index set $G$. Their operator algebra is generated by evaluations $\psi(g) = (\psi_1(g),\ldots, \psi_s(g))$ of an $s$-dimensional complex vector field $\psi$ over $G$, obeying the  
canonical anti-commutation relations
\begin{align*}
    &\big\{\psi_a(f),\psi_b(g)\big\}=\left\{\psi_a^\dagger(f),\psi_b^\dagger(g)\right\}=0\,,\\
    &\left\{\psi_a^\dagger(f),\psi_b(g)\right\}=\delta_{fg}\delta_{ab}\, \quad \forall\,f,g \in G,\ a,b,\in\{1,2,\ldots,s\}.
\end{align*}

By linearity, we mean the interaction among systems is described by linear transition functions $A_{gf}$, allowing us to write the evolution of system $\psi(g)$ as
\begin{equation}
\psi(g,t+1)=\alpha[\psi(g,t)]=\sum_{f \in S_g} A_{gf}\,\psi(f,t)\,, 
\end{equation}
with  $S_g \subset G$ the set of systems interacting with $\psi(g)$. Locality is phrased by imposing $S_g$ being finite for every $g$. Such a locality condition introduces a notion of causal cone in the emergent spacetime lattice. The homogeneity requirement amounts to ask that the cardinality of $S_g$, as well as the set of functions $\{A_{gf}\}_{f\in S_g}$, are independent of $g$; whence we can identify $S_g=S$ and the functions $A_{gf}=A_h$ for some $h\in S$. Moreover, we impose $A_{gf} \neq 0$ if and only if $A_{fg} \neq 0$. We are thus assuming that the QCA evolution does not discriminate systems. It follows that the structure of the connections between systems can be regarded as the application of generators (and their inverses) $h \in S $
of a discrete group---that we identify with $G$---which allows one to move from an element $g \in G$ to another element $f = hg \in G$. We can then define a spatial lattice as one possible Cayley graph of this group \cite{cayley}. The graph describes the interacting set of systems, by taking them as the nodes of the graph, with the links corresponding to their interactions. The isotropy condition stipulates that no preferential direction exists on the lattice. Mathematically, this necessitates the presence of a permutation group acting on $S$ which can be faithfully represented on the internal degrees of freedom. To confine our analysis to the dynamics on an emergent flat Minkowski spacetime, we assume the group $G$ to be virtually abelian, meaning $G$ contains an abelian subgroup of finite index. These are the groups whose Cayley graph can be quasi-isometrically embedded in Euclidean space \cite{de2000topics}. Finally, the unitarity assumption---namely reversibility of the algorithm---requires the local rule of a QCA to be a unitary representation $U_S$ of $\sum_{h \in S} A_{h}$. The quantum walk $W$ associated with the QCA $\alpha$, acting on the Hilbert space $\mathbb C ^s \otimes l^2(G)$, is obtained by 
considering the evolution of single-particle states
\begin{align*}
\varphi_a(g)=\psi^\dag_a(g)\ket\Omega,
\end{align*}
where $\ket\Omega$ denotes the \emph{vacuum} state vector. The quantum walk is given by
\begin{align*}
\varphi(g,t+1)=W\varphi(g,t+1)=\sum_{f \in S_g} A^*_{gf}\,\varphi(f,t)\,
\end{align*}
and thus it contains all the information that is needed in order to reconstruct $\alpha$~\cite{PhysRevA.90.062106}. 

All the QCAs with $s=2$, $G=\mathbb Z$ and $S=\{1\}$ of fermionic systems---whose internal degrees of freedom will be denoted as $R,L$---whose dynamics is linear in the field and fulfils the conditions above, are unitarily equivalent to the following QW on $\mathcal H = \mathbb C^2 \otimes l^2(\mathbb Z)$, which is called the Dirac QCA (or Quantum Walk) in one dimension:
\begin{equation}\label{W}
\begin{gathered}
    W=\begin{pmatrix}
        nS_{R} & im I \\
        im I & nS_{L}
    \end{pmatrix}\,,\quad n,m\in\mathbb{R}^+\,,\; n^2+m^2=1\,,\\
     \Psi(t+1)=W\Psi(t)\,, \quad  \Psi(t) =  \left(\ldots,\psi(x-1,t),\psi(x,t),\psi(x+1,t),\ldots\right)^\intercal\,,\\
     \psi(x,t)=\begin{pmatrix}
     \psi_R(x,t) \\
         \psi_L(x,t)
     \end{pmatrix}\,,
    \end{gathered}
\end{equation}
 where $I$, $S_R = \sum_{x \in \mathbb{Z}} \dyad{x+1}{x}$ and $S_{L}=S_R^\dagger$ are the identity, the right shift and the left shift operators on $l^2(\mathbb Z)$, respectively. The field component at time $t+1$ and at site $x$ depends only on the field components at sites $x \pm 1$ at time $t$ (first-neighboring scheme). Moreover, since $ W$ commutes with translations along the lattice, the automaton can be diagonalized in the wave-vector space. In wave-vector representation the Dirac QCA becomes
\begin{equation}\label{Wtilde}
    W =  \int_{B} dk\, \widetilde{W}(k)  \otimes \dyad{k}{k}\,, \quad   \widetilde{ W}(k) = \begin{pmatrix}
        ne^{-ik} & im \\
        im & ne^{ik}
    \end{pmatrix} \,,
\end{equation}
with $\dyad{k}{k} $ defined on $ L^2(B)$, where $B\coloneqq [-\pi,\pi]$ denotes the Brillouin zone.
Upon identifying the parameter $m$ with the mass and $k$ with the momentum of a particle, in the low-mass and low-momentum regime, it can be shown that the evolution described by the automaton cannot be discerned from the one prescribed by the Dirac equation of quantum field theory \cite{BISIO2015244}. Hence the name Dirac QCA.

\subsection{Path-sum solution}
The Dirac automaton in space representation can be conveniently expressed as \cite{dariano2014pathint}
\begin{equation}
\label{urlf}
    W=W_R\otimes S_R+W_L\otimes S_{L}+W_F\otimes I\,,
\end{equation}
along with the following binary encoding 
\begin{equation*}
    W_R \coloneqq nW_{00} \;\;\;,\;\;\;    W_L \coloneqq nW_{11} \;\;\;,\;\;\; W_F \coloneqq im(W_{01}+W_{10})\,,
\end{equation*}
where
\begin{gather*}
    W_{00}=\begin{pmatrix}
        1 & 0\\
        0 & 0
    \end{pmatrix}\,,\hspace{0.1\linewidth}  W_{11}=\begin{pmatrix}
        0 & 0\\
        0 & 1
    \end{pmatrix}\,,\\
    W_{01}=\begin{pmatrix}
        0 & 1\\
        0 & 0
    \end{pmatrix}\,,\hspace{0.1\linewidth}  W_{10}=\begin{pmatrix}
        0 & 0\\
        1 & 0
    \end{pmatrix}\,.
\end{gather*}
One can check that these binary matrices form a closed algebra under multiplication and satisfy a simple composition rule
\begin{equation}
\label {algebrau}
    W_{ab}W_{cd}=\frac{1+(-1)^{b\oplus c}}{2}W_{ad}=\delta_{bc}W_{ad}\,,
\end{equation}
where \(a\oplus b\coloneqq(a+b)\,\mbox{mod}2\) . Each dynamical step of the automaton consists of a shift $S_h$ acting on coordinates space according to the action of the corresponding transition matrix \(W_h\) on the internal degrees of freedom, with \(h\in\left\{R,L,F\right\}\) and \(S_F \equiv I\). A right shift (\(R\)) increases the site coordinate of right modes $\psi_R$ by one, a left shift (\(L\)) decreases the site coordinate of left modes $\psi_L$ by the same amount, and a free shift (\(F\)) leaves the site coordinate unchanged upon flipping right to left mode and vice-versa. In the one-particle sector, the QCA describes a Quantum Walk.
The evolution of the field over \(T\) steps is given by $W^T$ and can be thought of as a path connecting sites on the space-time lattice $\mathbb{Z} \times \mathbb{N}$.
We can then associate the generic path connecting \(x\) to \(y\) in \(T\) discrete time steps with a string \(s=s_1s_2\ldots s_T\in\left\{R,L,F\right\}^T\) of transitions, which gives the overall transition matrix \(\mathcal{W}(s)\)
\begin{equation*}
    \mathcal{W}(s) \coloneqq W_{s_T} W_{s_{T-1}}\cdots W_{s_1}\,.
\end{equation*}
The \emph{path sum} approach consists of expressing the state of the system at a given time as the outcome of all the possible evolutions allowed by the dynamics, each identifying a path on the space-time lattice. In this scenario, this translates into expressing the state \(\psi(x,t)\) as the result of the action of \(\mathcal{W}(s)\) on the generic initial state \(\psi(y,0)\), summed over all possible paths \(s\) and all points \(y\) in the past causal cone of site \((x,t)\):
\begin{equation}
\label{evolpsi}
    \psi(x,t)=\sum_y\sum_{s}\mathcal{W}(s)\,\psi(y,0)\,.
\end{equation}
Upon denoting with \(s^f\) the generic path containing \(f\) occurrences of the \(F\)-transition, eq. \eqref{evolpsi} becomes
\begin{equation}
\label{evolpsi2}
    \psi(x,t)=\sum_y\sum_{f=0}^{t-|x-y|}\sum_{s^f}\mathcal{W}\left(s^f\right)\,\psi(y,0)
\end{equation}
In a path \(s^f\), the \(F\) transitions identify \(f+1\) slots
\begin{equation*}
\tau_1 F \tau_2 F \cdots F \tau_{f+1}
\end{equation*}
where \(\tau_i\) denotes a (possibly empty) string of \(R\) or \(L\). The composition rule in eq. \eqref{algebrau} forbids the string \(s\) codifying a generic path from containing substrings of the form
\begin{align*}
s_{i}s_{i-1}&= RL   \;\;\;\;\; &s_{i}s_{i-1}&= LR \\
s_{i}s_{i-1}s_{i-2}&= RFR   \;\;\;\;\; &s_{i}s_{i-1}s_{i-2}&= LFL 
\end{align*}
as they give null transition amplitude. It follows that each sub-string \(\tau_i\) only consists of equal letters, namely either \(\tau_i=RR\cdots R\) or \(\tau_i=LL\cdots L\). Moreover, two consecutive strings \(\tau_i\) and \(\tau_{i+1}\) must contain different transitions. Therefore, all substrings \(\tau_{2i}\) occupying the even slots must be of one kind, and all substrings \(\tau_{2i+1}\) occupying the odd ones must be of the other kind. 
Exploiting the algebra \eqref{algebrau} and relying on combinatorial arguments, we can ultimately express the state of the field at \((x,t)\) as:
\begin{equation}\label{eq:solpathsum1p}
    \psi(x,t)=\sum_y\sum_{a,b\in\left\{0,1\right\}}\sum_{f=0}^{t-|x-y|}\,\left[ \alpha(f)\,c_{ab}(f)\,W_{ab}\right]\,\psi(y,0)\,,
\end{equation}
where the factor \(\alpha(f)\) is given by
\begin{equation*}
    \alpha(f)\coloneqq(im)^{f}\,n^{t-f}
\end{equation*}
with \(f\) the number of \(F\)-type transitions within the path connecting the field at $(y,0)$ to the field at $(x,t)$. The coefficients \(c_{ab}(f)\) account for the number of strings \(s^f\) corresponding to paths giving \(W_{ab}\) as the total transition matrix and can be computed by the following product of binomial coefficients
\begin{equation*}
    c_{ab}(f)=\binom{\mu_+-\nu}{\frac{f-1}{2}-\nu}\binom{\mu_-+\nu}{\frac{f-1}{2}+\nu}\,,
\end{equation*}
where
\begin{equation*}
    \nu=\frac{ab-\bar{a}\bar{b}}{2} \;\;\;\;\;\;\; \mu_{\pm}=\frac{t\pm(x-y)-1}{2}\,,
\end{equation*}
and satisfy \(c_{aa}(2k+1)=c_{a\bar{a}}(2k)=0\), with \(\bar{a}=a\oplus 1 \). The analytical solution of the Dirac walk (\ref{eq:solpathsum1p}) can also be expressed in terms of \textit{Jacobi polynomials} \(P_{k}^{(\alpha,\beta)}\) by explicitly computing the sum over \(f\) (see \cite{dariano2014pathint}). 

\section{Thirring QCA}\label{thirring}
To accomplish interacting quantum walkers, namely true QCAs, we must involve unitary transformations that are non-linear in the degrees of freedom \cite{centofanti2024}. Non-linearity naturally stems from the discreteness of the evolution. While a continuous framework might inherently imply a continuous evolution of the Hilbert space basis representing a local system, discrete steps in QCA framework lack an inherent method to compare the local basis at subsequent times. Consequently, we allow a free misalignment, introducing a local unitary evolution at each timestep - alongside the linear one - that is non-linear in the fermionic modes, while preserving topological symmetries.\\ 
We introduce the most general on-site number-preserving interaction in the one-dimensional Dirac QCA, particularly focusing on the two-particle sector. This type of interaction characterises a few notable integrable quantum systems \cite{PhysRevLett.20.1445, PhysRevD.11.2088, Korepin:1979qq,essler2005one} such as Hubbard’s \cite{doi:10.1098/rspa.1963.0204} and Thirring’s \cite{THIRRING195891} models. Hence, the present model is denoted as \textit{Thirring Quantum Cellular Automaton} \cite{bisio2018thirring}.
When discussing non-interacting particles, the one-particle sector completely specifies the dynamics. Indeed, the evolution of \(N\) free Dirac fermions is described by the operator \(W_N\coloneqq W^{\otimes N}\) acting on the Hilbert space \(\mathcal H^{\otimes N}\), obtained by the tensorisation of the single-particle evolution (\ref{W}). The Thirring QCA is defined as the most general on-site coupling $J(\chi)$ of one-dimensional many Dirac fermions, that preserves the number of particles. It is demonstrated \cite{PhysRevB.44.12413} that such interaction takes the form
\begin{equation}
\label{Jthirring}
J(\chi) \coloneqq e^{i\,\chi\, n_R(x)n_L(x)}
\end{equation}
where \(n_a(x)=\psi_a^\dagger(x)\psi_a(x)\) is the number operator at site \(x\) with internal state \(a\in\left\{R,L\right\}\), and \(\chi\in\left[-\pi,\pi\right]\) is the automaton coupling constant. The single-step evolution consisting of both the free evolution and the on-site interaction for $N$ particles, from now on referred as \textit{Thirring QCA}, is therefore defined by
  \begin{equation}
  \label{qcathirringn}
      U_N\coloneqq W_N\, J(\chi)\,.
  \end{equation}
Given that this QCA shares the same interaction term as the Hubbard and Thirring models, which are known to be integrable, it is natural to question its own integrability. All known quantum integrable systems are solved via the \textit{Bethe ansatz}. This method consists of solving the two-particle dynamics, formulating an ansatz for the solution of the N-particle case based on the two-particle solution, and verifying that the ansatz provides all the solutions. However, in the case of the Thirring QCA, the discrete nature of time evolution introduces nontrivial differences in the dynamics and results in a distinct phenomenology. Thus, the conventional \(N\)-particle ansatz cannot be straightforwardly applied to the Thirring QCA, leaving the question of its integrability unsolved.

We briefly recall here the solution in the two-particle sector (see Ref.\cite{bisio2018thirring}).\\ 
In the two-particle case, \(N=2\), it is convenient to work in the centre-of-mass basis \(\ket{a_1,a_2}\ket{y}\ket{w} \in \mathcal H^{\otimes 2}\), where \(a_i \in\left\{R,L\right\}\), while \(y=x_1-x_2\) and \(w=x_1+x_2\) denote the relative position and the centre-of-mass coordinates of the particles, respectively. The interaction operator can thus be represented as
\begin{equation*}
     J(\chi) = e^{i\,\chi\, \delta_{y,0}(1-\delta_{a_1,a_2})}\,.
\end{equation*}
Defining \(k,\,p \in B\) as the (half) relative momentum and the (half) total momentum of the particles respectively, the free QCA in the momentum representation is written as
\begin{gather*}
    W_2=\int_{B\times B} dk\;dp\;\widetilde{W}_2(k,p)\otimes\ket{k}\bra{k}\otimes\ket{p}\bra{p}\,, \\
    \widetilde{W}_2(k,p)\coloneqq\widetilde{W}(p+k)\otimes\widetilde{W}(p-k)\,,
\end{gather*}
with $\widetilde{W}$ as in (\ref{Wtilde}).
Since the interacting dynamics described by \(U_2\) commutes with the translations in the center-of-mass coordinate \(w\), it is convenient to write the automaton in the hybrid basis \(\ket{a_1,a_2}\ket{y}\ket{p}\)
\begin{gather*}
     U_2=\int_B dp\, U_2(\chi,p)\otimes\ket{p}\bra{p}\,, \quad U_2(\chi,p)\coloneqq W^h_2(p) {J}^h(\chi)\\
     W^h_2(p) \coloneqq mn\begin{pmatrix}
    \frac{n}{m}e^{-i2p} & ie^{-ip}T_y & ie^{-ip}T_y^\dagger & -\frac{m}{n} \\
    ie^{-ip}T_y & \frac{n}{m}T_y^2 & -\frac{m}{n} & ie^{ip}T_y^\dagger \\
    ie^{-ip}T_y^\dagger & -\frac{m}{n} & {\frac{n}{m}T_y^\dagger}^2 & ie^{ip}T_y^\dagger \\
     -\frac{m}{n} & ie^{ip}T_y & ie^{ip}T_y^\dagger & \frac{n}{m}e^{i2p}
    \end{pmatrix} \quad  {J}^h(\chi)\coloneqq \begin{pmatrix}
        I & 0 & 0 & 0 \\
        0 & e^{i\,\chi\, \delta_{y,0}}I & 0 & 0\\
        0 & 0 & e^{i\,\chi\, \delta_{y,0}}I & 0 \\
        0 & 0 & 0 & I \\    
    \end{pmatrix}
\end{gather*}
where the superscript $h$ stands for "hybrid", and $T_y,\,T_y^\dagger$ are the right and left shift in the relative coordinate $y$.
Solving the two-particle dynamics amounts to the diagonalisation of the infinite-dimensional operator \(U_2(\chi,p)\). Although, generally, it is not possible to obtain the analytical solution for an infinite-dimensional eigenvalue problem, the authors of \cite{bisio2018thirring} show that the eigenvalue problem 
\begin{equation*}
U_2(\chi,p)\ket{\psi}=e^{i\omega}\ket{\psi}    
\end{equation*}
can be suitably reduced to a finite set of algebraic equations that admit an analytical solution. This is done by considering the linear difference equation
\begin{align*}
    U_2(\chi,p)&\mathbf{f}_{p,\omega,\chi}=e^{i\omega}\mathbf{f}_{p,\omega,\chi}\\
    &\mathbf{f}_{p,\omega,\chi}\; :\; \mathbb{Z} \to \mathbb{C}^4 \;\;\;,\;\;\;\omega\in\mathbb{C}
\end{align*}
for any possible value of \(\chi\) and \(p\).  Since, in the new coordinates, the interacting term acts only at the origin, for \(y > 0\) such an equation results in a linear recurrence relation with constant coefficients, which are then determined by requiring the solutions to be anti-symmetric eigenvectors (or generalised eigenvectors) of \(U_2(\chi,p)\) intended as an operator on the Hilbert space \(\mathbb{C}^4\otimes l^2(\mathbb{Z})\). The solutions thus obtained present some rather peculiar features, most notably:
\begin{itemize}
    \item there are bound states for every value of the total momentum, which also exist in the free case;
    \item there are four classes of scattering solutions, as opposed to the two obtained with the usual quantum field theory approach;
    \item the Thirring QCA in the continuum limit does not straightforwardly yield the Thirring model.
\end{itemize}
\section{Preliminary results}\label{mathtools}

\begin{figure}[h]
    \centering
    \includegraphics[width=0.85\textwidth]{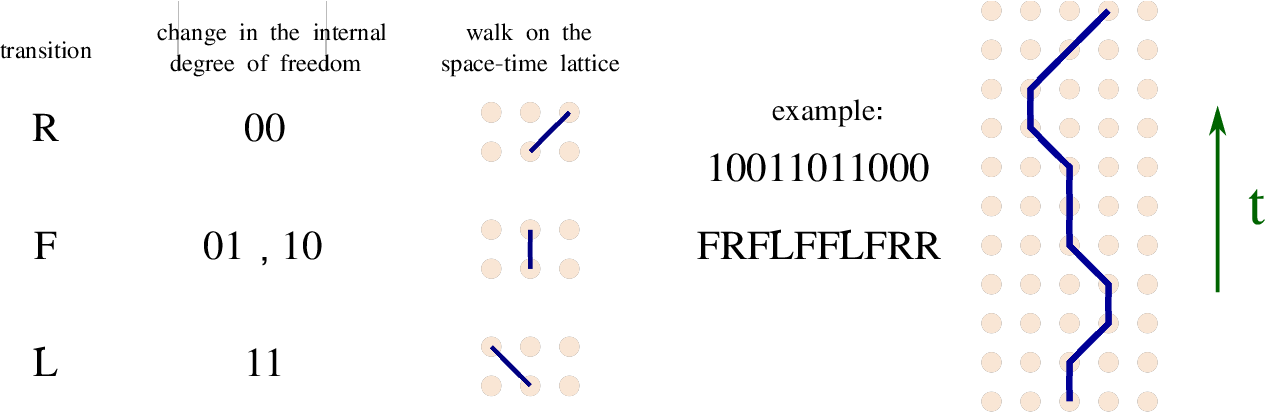}
    \caption{On the left, the correspondence between different ways of representing the evolution of the particle at each step of the automaton is given. On the right, an example of the three representations for a generic walk is shown. The strings, whether representing lattice transitions or internal degrees of freedom, are ordered with time progressing from left to right, while on the lattice, time flows from bottom to top.}
    \label{fig:rapp}
\end{figure}

In this section, we present the mathematical tools useful for analyzing and classifying particle paths in the interacting scenario. All detailed proofs can be found in Appendix \ref{app}. The central result shown here is the fact that, given the initial conditions of the field, the free dynamics of the automaton univocally determines its state at each site of the space-time lattice. Combined with Pauli's exclusion principles, this establishes necessary boundary conditions for particles to interact at any space-time point. It is convenient to denote by bit $0$ the right mode of the field and by bit $1$ the left mode. The evolution of one particle can thus be described both with a string of transitions, made up of \(R,L,F\), or with a binary string associated with the state of the particle at each time step. The main advantage of employing binary strings lies in their lack of constraints regarding the arrangement of their elements. Unlike translations, where substrings $LR$ and $RL$ are prohibited by the algebra of transition matrices, any sequence of $0$s and $1$s constitutes a valid path. We begin by defining a mapping between binary strings of internal degrees of freedom and the corresponding strings of translations on the lattice, which serves to elucidate the argument  (see Fig.\ref{fig:rapp}).

\begin{Definition} \label{defM} Let \(\mathcal{M}\) be a map associating a pair of bits, encoding the state of the particle for two subsequent evolution steps, with the corresponding translation on the lattice
\begin{align*}
    \mathcal{M} \;\; : \;\; \left\{0,1\right\}^{2} &\to \left\{R,L,F\right\}\\
    b_1b_2 &\mapsto \mathcal{M}_{b_1b_2}
\end{align*}
according to the following rules:
\begin{equation*}
    \mathcal{M}_{00}=R \;\;\;;\;\;\; \mathcal{M}_{11}=L \;\;\;;\;\;\; \mathcal{M}_{01}=\mathcal{M}_{10}=F\,.
\end{equation*}
A correspondence between the binary string of subsequent internal degrees of freedom  \(b=b_0\cdots b_{T}\) and the string of translations \(s=s_1\cdots s_T\) can be established:
\begin{align*}
    \overline{\mathcal{M}} \;\; : \;\; \left\{0,1\right\}^{T+1} &\to \left\{R,L,F\right\}^T\\
    b &\mapsto \overline{\mathcal{M}}(b)=s
\end{align*}
with
\[s_k=\mathcal{M}_{b_{k-1}b_k} \;\;\;\;\; 1\le k \le T\,.\]
\end{Definition}
We introduce a function that takes a single-step transition as an argument and outputs the corresponding displacement relative to the position of the particle at the previous step. Then, given a string of translations, the resulting displacement between the starting and ending points can be computed as the sum of the displacements relative to each of its elements. This allows us to define the set $S_{x_{in}\to x_{out}}^T$ $\left(B_{x_{in}\to x_{out}}^T\right)$ containing the strings of translations (bits) that represent a path connecting \(x_{in}\) to \(x_{out}\) in \(T\) time steps.
\begin{Definition} 
The sets of strings of translations and strings of bits representing a path connecting \(x_{in}\) to \(x_{out}\) in \(T\) time steps are respectively given by
 \begin{align*}
S_{x_{in}\to x_{out}}^T \coloneqq&\left\{ s=s_1\cdots s_T\in\left\{R,L,F\right\}^{T} \;\mbox{s.t.} \; \sum_{i=1}^T \Delta(s_i)=x_{out}-x_{in} \right\}\,,
\end{align*}
\begin{align*}
    B_{x_{in}\to x_{out}}^T \coloneqq \left\{ b=b_0\cdots b_T\in\left\{0,1\right\}^{T+1} \;\mbox{s.t.} \; \overline{\mathcal{M}}(b)=s\in S_{x_{in}\to x_{out}}^T\right\}\,,
\end{align*}
with \(\Delta\) the \textit{displacement operator} 
\begin{align*}
    \Delta \;\; : \;\; \left\{R,L,F\right\} &\to \left\{0,\pm 1\right\}\\
    \mathcal{M}_{ab} &\mapsto \Delta(\mathcal{M}_{ab})
\end{align*}
such that 
\begin{equation*}
    \Delta(\mathcal{M}_{ab}) =(-1)^{ab}\left[1-(a\oplus b)\right]\,.
\end{equation*}
\end{Definition}
We proved that any permutation of the internal bits of a string does not change the extremal points of the associated path. Thus, the number of \(0\)s and \(1\)s contained in a binary string is related to the space-time coordinates of the initial and final sites of the path it describes. This relationship is made explicit in the following Lemma \ref{lemmaweight}, which is fundamental for the derivation of all subsequent results. 
\begin{Lemma}\label{lemmaperm} 
For any permutation \(\Pi\) of the internal bits \(\left\{b_i\right\}_{i=1}^{T-1}\) of a string \(b\in B_{x_{in}\to x_{out}}^T\) then \(\Pi(b)\in B_{x_{in}\to x_{out}}^T\). 
\end{Lemma}

\begin{Lemma}\label{lemmaweight} 
The number of ones contained in the string $ b=b_0\cdots b_T\in\left\{0,1\right\}^{T+1}$ is called the weight \(w(b)\) of the string $b$ 
\begin{align*}
w:\left\{0,1\right\}^{T+1}\to\mathbb{N}\quad\quad b \mapsto w(b)\coloneqq\sum_{i=0}^T b_i\,.
\end{align*} 
If $b \in  B_{x_{in}\to x_{out}}^T $, then
\begin{equation}
\label{pesiW}
      w(b)=\frac{1}{2} \left(T+(x_{in}-x_{out})+b_0+b_T\right)\,.
\end{equation}
\end{Lemma}
Eq. \eqref{pesiW} highlights the interplay between the internal degrees of freedom of the particle and its space-time coordinates at any given point in space and time, given its boundary conditions. \begin{Lemma}\label{coruni}
Given the initial conditions of the particle, i.e. space coordinate and internal degree of freedom at $t$, the internal degree of freedom at each site of the space-time lattice for $t\neq 0$ is univocally determined.
\end{Lemma}
This result, along with lemma \ref{lemmaweight}, implies that all binary strings connecting \(x_{in}\) to \(x_{out}\) in \(T\) steps have the same weight, namely
\begin{equation*}
    w(b)=w(b') \;\;\;\;\;\forall b,b' \in B_{x_{in}\to x_{out}}^T\,.
\end{equation*}

\begin{figure}[h]
    \centering
    \includegraphics[width=0.85\textwidth]{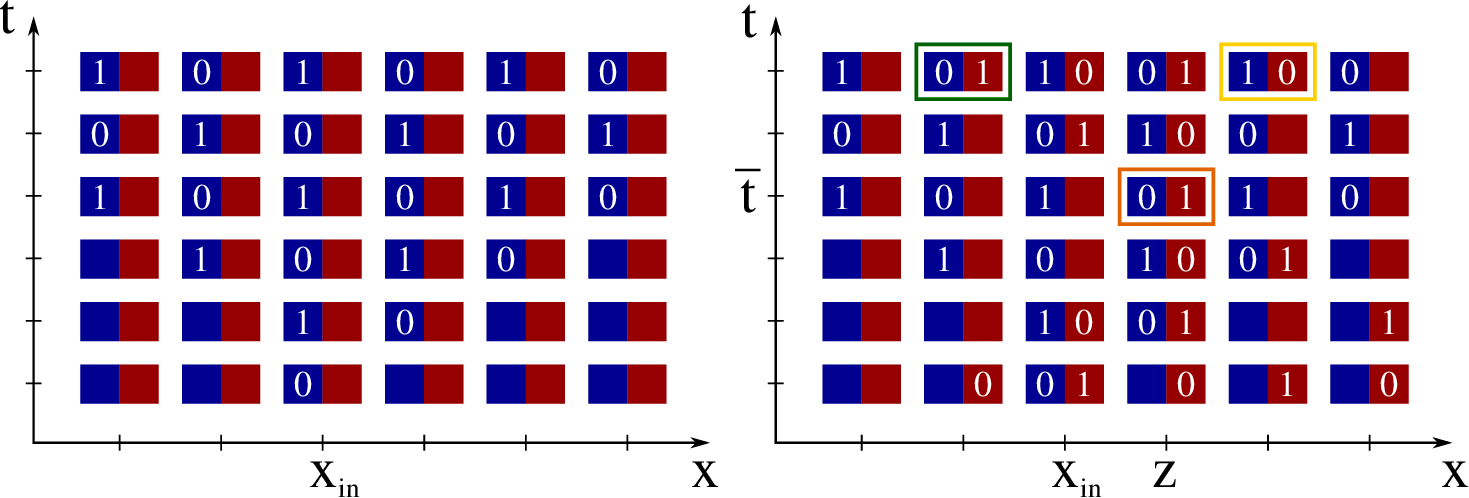}
    \caption{Sites of the space-time lattice can be represented as a pair of containers, each accomodating the internal degree of freedom of a particle. Particle $A$ is associated with red boxes, while particle $B$ with blue boxes. The left panel represents the internal state of particle $B$ at each time step upon choosing its initial state. On the right, observe that assuming to fix the interaction site \((z,\bar{t})\) (highlighted in orange), then, at the bottom row, equal (opposite) internal states are separated by an even (odd) number of sites. Secondly, suppose the final sites are those highlighted in green and yellow. Then there is no ambiguity in discerning the two particles based on their internal state, meaning if we measure the internal state at the green (yellow) site and read \(0\) we know it is the state of the particle evolving in the blue (red) containers, and vice-versa if we read \(1\). Thus the particles $A$ and $B$ are distinguishable.}
    \label{fig:tab}
\end{figure}

Let us now introduce a second particle. 
In the interacting scenario, a site can be thought of as a couple of identical containers, one for each particle, and the space-time lattice as a collection of such containers. Each container is assigned a bit \(0\) or \(1\) denoting the internal degree of freedom of the related particle. 
Upon fixing the mode of the particle, whose evolution is represented e.g. in the left containers, at a given site, one can fill each left container across the space-time lattice according to the rules prescribed by the free theory (summarised in Fig. \ref{fig:rapp}). For instance, by choosing the initial state for the particle to be \(0\), we obtain the configuration depicted in Fig. \ref{fig:tab} on the left. Thus, the free dynamics of the QCA univocally determines the internal degrees of freedom of the particles at every space-time site. Furthermore, since we are considering particles with fermionic statistics, they obey Pauli's exclusion principle. Regarding the present QCA, this translates into the requirement that no more than two particles can share the same space-time coordinate, and whenever two particles happen to be on the same site their internal degrees of freedom must be opposite.
To keep track of the relative position of the two particles along their evolution, we introduce the following definition.
\begin{Definition}
Let \((x,t)\) and \((y,t)\) be the space-time coordinates of particles \(A\) and \(B\) at time step \(t\), respectively. Their relative position at step \(t\) is then defined as \(\delta_t\coloneqq x-y\).
\end{Definition}

Due to the nature of the Thirring interaction (\ref{Jthirring}), necessary conditions for two particles to interact at step \(t\) are occupying the same site, i.e. \(\delta_t=0\), and possessing opposite internal degrees of freedom at that time. This observation together with Eq. \eqref{pesiW}, leads to a necessary condition on the initial states for an interaction to occur at a given point. For instance, consider the scenario represented in Fig. \ref{fig:tab} on the right, and suppose the particles interact at a given site \((z, \bar{t})\). By applying the usual rules for the free evolution (which is unitary and thus reversible), we can proceed backward in time and fill the right containers across the lattice with the internal state of the second particle. Notice that, at the initial time step (corresponding to the bottom row), whenever the second particle is in the same (opposite) internal state as the first one, they are separated by an odd (even) number of sites. 
\begin{Lemma}\label{lemcnec}
Consider two particle paths whose starting points at time $t=0$ are separated by \(\delta_0\) lattice sites. The paths are allowed to intersect at time $t\neq 0$ iff their internal degrees of freedom at $t=0$ are
\begin{itemize}
    \item opposite if \(\delta_0\) is even,
    \item equal if \(\delta_0\) is odd.
\end{itemize}
\end{Lemma}
As a consequence of Lemma \ref{lemcnec}, we show that, whenever two particles can interact, the particles are distinguishable, meaning that the final state of one particle cannot be the final state of the other. Indeed, let us consider once again the scenario illustrated in Figure \ref{fig:tab} on the right. Upon examination, it is immediate to notice that if particle \(A\) ends up at the generic site \((x,t)\) with an internal degree of freedom \(b_t\), then particle \(B\) must be in state \(\bar{b_t}\) whenever it lands on that same site. This allows us to distinguish the two particles based on their final internal degree of freedom.
\begin{Corollary}\label{cor}
Two particles satisfying the necessary condition for an interaction to occur are distinguishable. 
\end{Corollary}
To conclude this section, one last result is showcased, which will be useful in classifying paths of interacting particles based on their reciprocal position at the beginning and the end of a process. 
\begin{Lemma}\label{lemmadist} 
Following an interaction, the relative position of the particles involved changes sign, i.e. if an interaction occurs at step \(t\), then 
\begin{equation}
    \label{lemin}
    \delta_{t-1}\delta_{t+1}\le 0\,.
\end{equation}
\end{Lemma}
Upon fixing both the initial and final conditions for an interaction process, Lemma \ref{lemmadist} allows us to establish if an even or odd number of interactions occurred.
\begin{Corollary}
Let \(x_{in}\), \(x_{out}\), \(y_{in}\), \(y_{out}\), be the boundary space coordinates of two particles, with \(x_{in} < y_{in}\). Let $k \in \mathbb{N}_0$, then
\begin{itemize}
    \item  if \(x_{out} < y_{out}\)  the particles interact $2k$ times (they may not interact at all),
\item if \(x_{out} > y_{out}\)  the particles interact $2k+1$ times (they must interact at least once).
\end{itemize}
  
\end{Corollary}

\section{Perturbative approach in the number of interactions}\label{pert_int}

We now turn to the description of the perturabtive techniques developed for the evaluation of transition matrices in the two-particle sector of the Thirring QCA. The path sum approach requires a thorough classification of all the allowed pairs of paths based on the number of intersections between them, to determine how many times the on-site coupling \(J(\chi)\) must be accounted for. The computational complexity of this task is exponential in the size of space-time lattice. To reduce complexity we adopted the following expedient, allowing for the separation of the notions of intersection between paths and interaction of particles. We write the QCA for $T$ evolution steps of two particles by expressing each interaction term as \(J=I+(J-I)\), where \(I\) denotes the identity on the Hilbert space $\mathcal{H}^{\otimes 2}$
\begin{align*}
    U_2^T&=JW_2JW_2\cdots JW_2=[I+(J-I)]W_2[I+(J-I)]W_2\cdots [I+(J-I)]W_2\,.\\
\end{align*}
We then group terms based on the number of interactions within them, thus obtaining 
\begin{equation*}
    U_2^{T} = W_2^T+\sum_{k=1}^T \mathcal U^{(k)}
\end{equation*}
where \(W_2^T\) corresponds to the free evolution, while \(\mathcal{U}^{(k)}\) denotes the contribution of paths containing \(k\) interactions
\begin{equation*}
    \mathcal{U}^{(k)}\coloneqq\sum_{\overline{\alpha}\in A_k}W_2^{\alpha_0}(J-I)W_2^{\alpha_1}(J-I)\cdots W_2^{\alpha_{k-1}}(J-I)W_2^{\alpha_{k}}
\end{equation*}
with
\begin{equation*}
    A_k\coloneqq \left\{\overline{\alpha}=(\alpha_0,\ldots,\alpha_{k})\in\mathbb{N}^{k+1} \;\;\; \mbox{s.t.} \;\; \sum_{i=0}^{k}\alpha_i=T \;\;\; \wedge\;\; \alpha_i\neq 0 \;\; \forall \; i>0 \right\}\,.
\end{equation*}

\begin{figure}[h]
    \centering
    \includegraphics[width=0.45\textwidth]{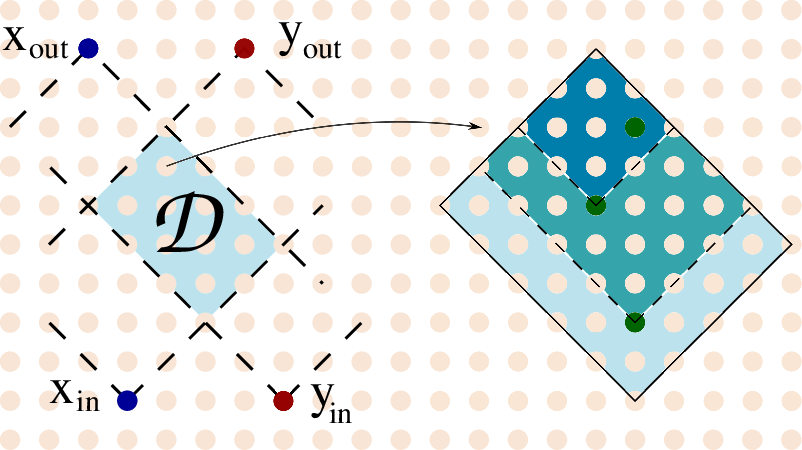}
    \caption{On the left, a graphical representation of the intersection $\mathcal{D}$ between the future causal cones of the initial sites and the past causal cones of the final sites. In such a region, paths can intersect and thus interact. An (enlarged) instance of the nested disposition of three causally connected vertices is presented on the right.}
    \label{fig:d}
\end{figure}

We remark that within this context, the term "perturbative" takes on a slightly different meaning than the usual one. We are power expanding in the number of interactions, but we are not claiming that some contributions are negligible compared to others. In this sense, the result of the calculation is exact. To evaluate the contribution of \(\mathcal{U}^{(k)}\) to the total transition matrix we proceed as follows:
\begin{enumerate}
    \item fix a priori \(k\) time steps where the interactions occur;
    \item evaluate the transition matrix associated with paths interacting at the chosen steps;
    \item sum over all possible ways of choosing such time steps according to the causal structure of the lattice (see Fig. \ref{fig:d}). 
\end{enumerate}
Let us stress that \(\mathcal{U}^{(k)}\) describes the evolution of paths which, in general, intersect a number of times greater than \(k\). However, if the intersection occurs at a different time from those specified in item \(1\) of the procedure, it is not considered an interaction. In a way, one can think of this scenario as if the experimenter had the ability to turn the interaction on or off at will. Then physical predictions comes from taking into account all the possible ways one can perform this operation (item 3). As a consequence, the result derived in Lemma \ref{lemmadist} no longer holds: the two particles must be treated as \emph{indistinguishable}, thus requiring an anti-symmetrization procedure at each interaction site. 

\begin{figure}[h]
    \centering
    \includegraphics[width=0.85\textwidth]{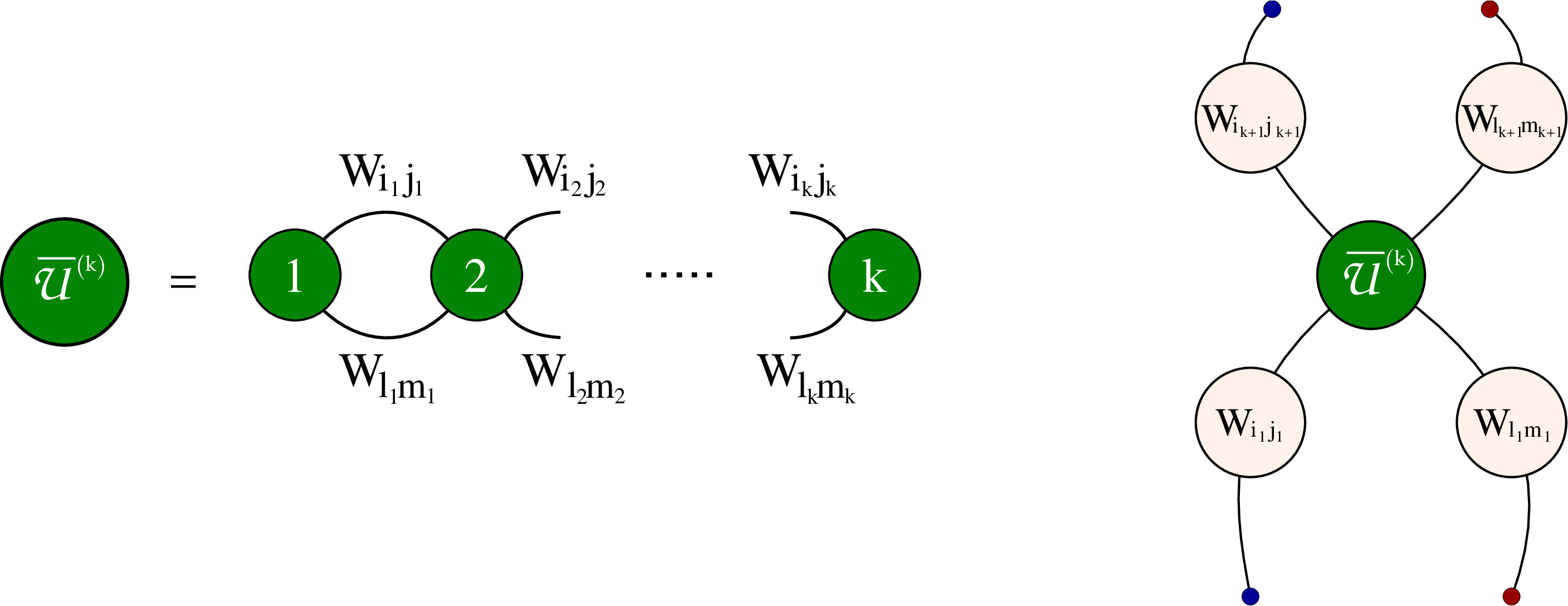}
    \caption{On the left, a diagrammatic representation of the matrix \(\overline{\mathcal{U}}^{(k)}\) describing the evolution of the particles from the first interaction to the last, within a \(k\)-interactions process, regardless of the boundary conditions. This matrix is subsequently contracted with the free transition matrices connecting to the initial and final states, yielding the \emph{total} transition matrix \( \mathcal{U}^{(k)}\) for the \(k\)-interactions process, depicted on the right.}
    \label{fig:amp}
\end{figure}

Upon fixing the number \(k\) of interactions, the particles evolve freely between interaction sites. To carry out step \(2\) of our procedure, we first compute the transition matrix describing the evolution of two particles from their first interaction to their last, temporarily setting aside considerations regarding boundary conditions (see Fig. \ref{fig:amp}). Since the transition matrix for the free dynamics is a linear combination of the four binary matrices \(W_{ij}\), \(i,j=0,1\), supposing \(k\) interaction sites and upon anti-symmetrization, then the amputated transition matrix for the interacting case is a linear combination of transition matrices of the form:
\begin{align*}
&\frac{1}{2^{k}}(I-E)(J-I)(I-E)(W_{i_1 j_1}\otimes W_{l_1 m_1})(I-E)(J-I)(I-E)\cdots \\
& \;\;\;\;\;\;\; \cdots (I-E)(J-I)(I-E)(W_{i_{k-1} j_{k-1}}\otimes W_{l_{n-1} m_{k-1}})(I-E)(J-I)(I-E)\\
=&\frac{1}{2^{k}}\,(e^{i\chi}-1)^{k}\,\overline{\mathcal{U}}^{(k)}(i_1,j_1,\ldots,i_{k-1},j_{k-1},l_1,m_1,\ldots,l_{k-1},m_{k-1})
\end{align*}
with \(   E \coloneqq \frac{1}{2}\sum_{i=0}^3 \sigma_i\otimes \sigma_i \) the exchange operator, where \(\sigma_0=I\) and \(\sigma_i\) (\(i=1,2,3\)) are the Pauli matrices, and
\begin{equation*}
    \overline{\mathcal{U}}^{(k)}\coloneqq (I-E)(J-I)(I-E)\prod_{n=1}^{k-1}\left[(W_{i_n j_n}\otimes W_{l_n m_n})(I-E)(J-I)(I-E)\right]\,.
\end{equation*}

The matrix \(\overline{\mathcal{U}}^{(k)}\) can be expressed more compactly by exploiting the algebra of the binary matrices \(W_{ij}\), resulting in a product of Kronecker \(\delta\)s encoding compatibility conditions for the internal degrees of freedom at each interaction site. To conclude step \(2\), we take into account boundary conditions and evaluate the multiplicity of paths contributing to the matrix \(\overline{\mathcal{U}}^{(k)}\) for the specific choice of time steps we made in step \(1\) of the procedure. The routine can be summarised as follows:
\begin{itemize}
    \item  divide the binary strings associated with the particles' evolution into \(k+1\) sub-strings, with \(k\) the (fixed) number of interactions; the cuts are made in correspondence with the time steps chosen in step \(1\)
    \item re-distribute the weight of each of the two string into its sub-strings.\footnote{This is made possible by Lemma \ref{lemmaperm}, as long as the first and last bit are keep fixed, which depend on the boundary conditions}
\end{itemize}
Finally, we move on to step \(3\) and sum over all possible ways of choosing the time steps in step \(1\). By iterating this procedure, we inferred the general structure of the transition matrix \(\mathcal{U}^{(k)}\) for an arbitrary number \(k\) of interactions. It constitutes of:
\begin{itemize}
    \item \(4(k+1)\) sums over the indices associated with the transition matrix describing the free evolution from the initial sites to the first interaction, from each interaction site to the following one and from the last interaction to the final sites
    \item \(2k\) sums over the space-time coordinates of each interaction site
    \item the product of \(2(k+1)\) Jacobi polynomials 
    \item the product of \((e^{i\chi}-1)^k/2^k\) with the matrix \((W_{i_1 j_1}\otimes W_{l_1 m_1})\overline{\mathcal{U}}^{(k)}(W_{i_{k+1}j_{k+1}}\otimes W_{l_{k+1} m_{k+1}})\) 
\end{itemize}

In conclusion, our endeavors were directed towards formulating a set of rules that would facilitate the calculation of transition amplitudes for interaction processes, analogous to the application of Feynman rules in quantum field theory.
However, our approach was complicated by an additional challenge proper of the QCA framework. While the canonical path integral formulation of QFT prescribes all paths, even non-physical ones - once weighted by the exponential of the Euclidean action - to be taken into account when evaluating transition amplitudes, our Hamiltonian-free approach does not enjoy this freedom. Restricting to physical paths requires identifying all allowed space coordinates for each interaction site so that they are causally connected (see Fig. \ref{fig:d}). This culminates in an iterative procedure analogous to a search problem, which is notoriously demanding in computational terms. Consequently, the algorithm for the perturbative method based on the number of interactions, despite characterizing pertinent interacting paths, offers no substantial advantage over a naive classical simulation of the automaton in predicting physical quantities.

\section{Perturbative approach in the mass parameter}\label{pert_mass}

Owing to the notable challenges encountered in developing a perturbative approach in the number of interactions, we focused on identifying how the parameters of the automaton influence its behaviour, to discern physical regimes characterised by reduced dynamical complexity. Within the path sum framework, this feature is embodied in the regularity of the walks, where particles tend to either consistently shift or remain stationary. Consequently, the number of paths leading to an interaction diminishes, enabling their manual enumeration.

Upon identifying the parameter \(m\) of the Dirac QCA \(W\) (\eqref{W}) as the mass of the particle evolving on the space-time lattice, we can detach two different dynamical regimes, depending on its magnitude:
\begin{itemize}
    \item  if \(m\ll 1\), the off-diagonal terms in \(W\) contribute minimally to the dynamics, resulting in paths where the particle shifts at almost every step;
    \item if \(m\approx 1\), the contribution of the off-diagonal terms dominates the dynamics, resulting in paths where the particle rarely shifts.
\end{itemize}
In order to exploit this feature, it is convenient to describe the evolution in terms of the string of transitions: light particles are associated with strings containing very few \(F\)-type transitions, while heavy particles are associated with strings made up almost entirely of them. Therefore, the perturbative approach in the mass parameter translates into a perturbative approach in the number \(f\) of \(F\)-type transitions.

We begin by computing the allowed values of \(f\) for paths contributing to a process with given boundary conditions. This will provide us with a criterion to divide all admissible diagrams into classes of paths belonging to the same perturbative order.  Lemma \ref{coruni} states that, once the internal degree of freedom at a certain point in time and space is fixed, it is univocally determined at each other site of the space-time lattice. Consequently, the parity of \(f\) must be the same for all paths sharing the same boundary conditions. This stems directly from the fact that an \(F\)-type transition corresponds, in the binary representation, to either \(01\) or \(10\). Thus, \(f\) can be interpreted as the number of changes in the internal degree of freedom of the particle along its path. Therefore, if a particle departs from site \(x_{in}\) with internal state \(b_0\) and arrives at site \(x_{out}\) with state \(b_T=b_0\) (\(\overline{b_0}\)), an even (odd) number of changes must have occurred, meaning \(f\) must be even (odd).  We recall that the maximum value of \(f\)  for a path connecting \(x_{in}\) to \(x_{out}\) over \(T\) discrete-time steps is given by \cite{BISIO2015244}
\begin{equation*}
    f_{max}=T-|x_{in}-x_{out}|\,.
\end{equation*}
Then, the set of the admissible values of \(f\) for a path with given boundary conditions can be characterised as:
\[F(x_{in},x_{out},T)\coloneqq\left\{ f\in\mathbb{N}-\left\{0\right\} \;\;\;|\;\;\; f=f_{max}-2n \;\; n\in\mathbb{N} \right\}\]
assuming  \(f=0\) whenever \(f_{max}=0\), corresponding to a light-like path.

Since we are interested in classifying diagrams describing the evolution of two particles, which we assume have the same mass, henceforth \(f\) will denote the \emph{total} number of \(F\)-type transitions within the diagram, namely the sum of those of both particles. Then, the class of diagrams describing a process over $T$ steps with initial conditions \(x_{in}\),\(y_{in}\), final conditions \(x_{out}\), \(y_{out}\)  and \(f=n\) can be defined as follows:
\begin{align*}
    \mathcal{F}_{x_{in},x_{out}}^{y_{in},y_{out}}(n,T)\coloneqq&\left\{(a,b)\in F(x_{in},x_{out},T)\times F(y_{in},y_{out},T) \right.\\
    &\left. \mbox{s.t.} \;\;\; a+b=n\right\}
\end{align*}
We then introduce a finer classification based on how the total \(F\)-type transitions are distributed between the two particles. Namely, within the class of representative \(f\), we further divide into sub-classes \((f_1,f_2)\) such that \(f_1+f_2=f\). 

In our classification, we only considered diagrams involving at least one interaction. In the following, the path associated with the particle starting from the site with the lower (higher) space coordinate is represented in blue (red) and its extremal points are denoted with \(x_{in}\) and \(x_{out}\) (\(y_{in}\) e \(y_{out}\)).  Dotted lines represent alternative paths, while green lines correspond to paths both particles share. For convenience, topologically equivalent diagrams are grouped together, since they share the same combinatorial analysis, thereby resulting in transition matrices with the same coefficients. Such diagrams can be obtained from each other through the following reflections about the vertical and horizontal axis of the plane.
\vspace{0.3cm}

\begin{minipage}{\textwidth} 
\begin{minipage}{0.4\textwidth}
\begin{minipage}{\textwidth} 
\begin{align*}
    \mathcal{R}_v(W_{ij}\otimes W_{lm})&=W_{\bar{l}\bar{m}}\otimes W_{\bar{i}\bar{j}}\\
      \mathcal{R}_o(W_{ij}\otimes W_{lm})&=W_{\bar{m}\bar{l}}\otimes W_{\bar{j}\bar{i}}
\end{align*}
\end{minipage}   
\vspace{0.35cm}

\begin{minipage}{\textwidth} 
\centering
\includegraphics[width=0.85\textwidth]{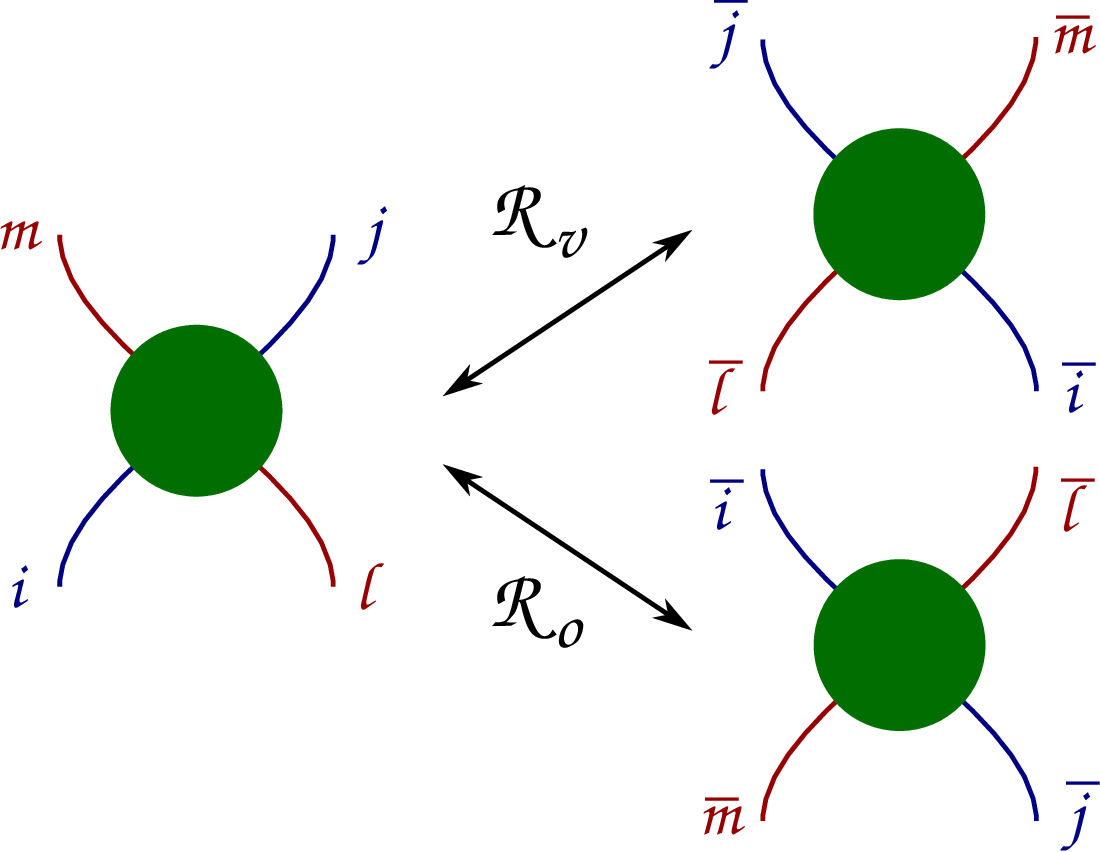} 
\end{minipage}

\end{minipage}
\hspace{0.5cm}
\begin{minipage}{0.4\textwidth}
\begin{minipage}{\textwidth} 
\begin{align*}
    \mathcal{R}_v(W_{ij}\otimes W_{lm})&=W_{\bar{l}\bar{m}}\otimes W_{\bar{i}\bar{j}}\\
      \mathcal{R}_o(W_{ij}\otimes W_{lm})&=W_{\bar{j}\bar{i}}\otimes W_{\bar{m}\bar{l}}
\end{align*}
\end{minipage}    
\vspace{0.35cm}

\begin{minipage}{\textwidth} 
\centering
\includegraphics[width=0.85\textwidth]{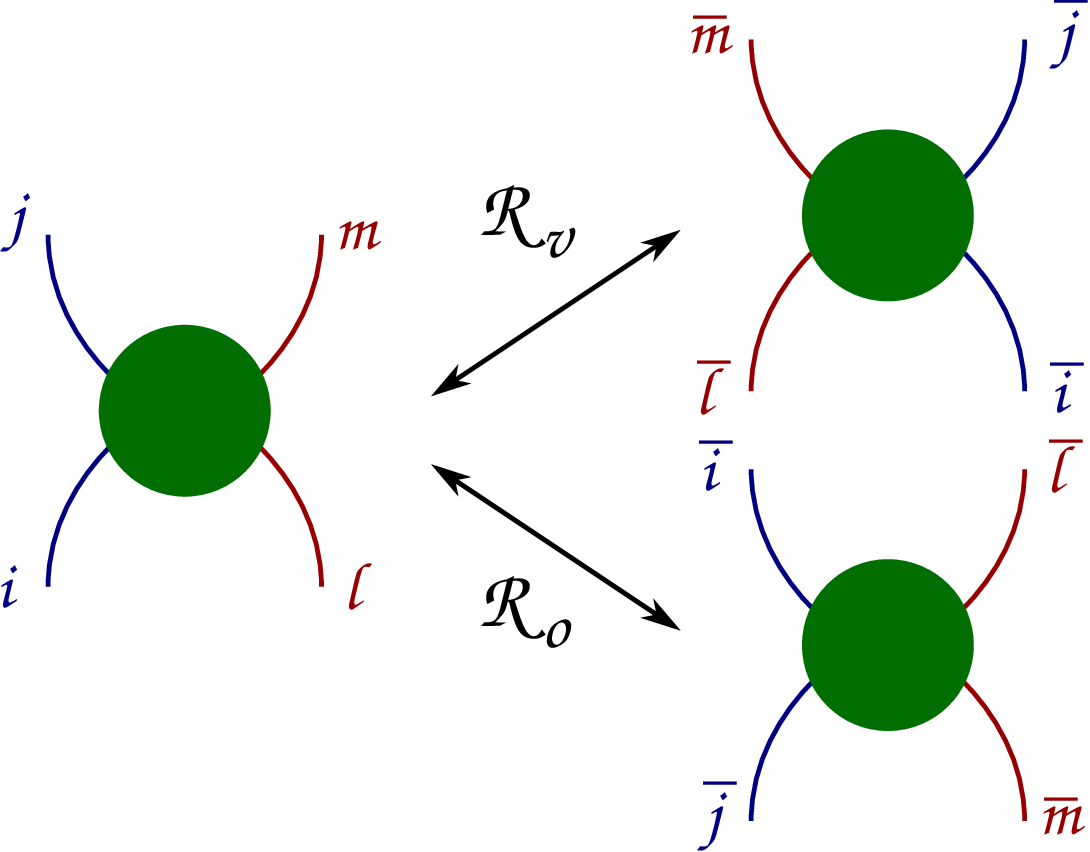} 
\end{minipage}
\end{minipage}
\end{minipage}

\vspace{0.3cm}
Lemma \ref{lemmadist} ensures that in the case shown on the left, when the relative position of the two particles changes sign after the interaction, they must interact an odd number of times. Conversely, in the case shown on the right, when particles' relative position does not change, they must interact an even number of times.

\subsection{Low-mass limit}
We begin by discussing the case of light particles, whose paths contain very few \(F\)-type transitions. We considered cases \(f=0,1,2,3\). For each of them, we classified all admissible diagrams containing at least one interaction and computed their transition matrix.

Let us begin by considering the trivial case \(f=0\), which only contains the sub-class \((0,0)\). From the algebra of the transition matrices \(W_h\), \(h=L,R,F\), it follows that sub-strings \(LR\) and \(RL\) are forbidden within the string of transitions, as they give null transition amplitude. Thus, in order to be able to change the direction of its path, a particle must make an \(F\)-type transition first. Therefore, diagrams with \(f=f_1+f_2=0\) correspond to a scenario where both particles evolve along light-like paths, since they must shift in the same direction at each time step, according to their initial state. Therefore, as long as their initial conditions allow it, they can only interact once. The resulting transition matrix is:
\[e^{i\chi}n^{2T}(W_{00}\otimes W_{11})\]
Notice that, whenever one of the two paths in the diagram is light-like, there can be at most one interaction. Indeed, the only way for them to interact more than once would be to shift in the same direction immediately after an interaction. However, there is no way for this to happen while obeying both Pauli's exclusion principle and the automaton's dynamics. Thus, for any value of \(f\ge 1\), any interacting diagram in the sub-class \((0,f)\) has transition matrix
\begin{equation*}
    e^{i\chi}n^{2T-f}(im)^fc_{a\;a\oplus f}(f)\begin{cases}
        W_{00}\otimes W_{a\;a\oplus f}\\
        W_{a\; a\oplus f}\otimes W_{11}
    \end{cases} \;\;\; a\in\left\{0,1\right\}
\end{equation*}
depending on which particle evolves along the light-like path. 

For sub-classes \((1,f-1)\) with \(f\ge2)\), the classification becomes, in general, more challenging. First of all, since neither path is light-like, it is possible to have more than one interaction. However, notice that the path with a single \(F\)-type transition can be considered as two light-like paths connected by it. Since each light-like path can accommodate at most one interaction, such diagrams can contain at most two interactions, one before said \(F\)-type transition, and one after. Therefore all diagrams in this subclass contain \(0\le n\le 2\) interactions. As a consequence of Lemma \ref{lemmadist}, the parity of the number of interactions depends on the relative position of the initial and final sites. Thus, when the relative position changes (which amounts to an odd number of interactions) the particles must interact once, and the resulting transition matrix is
\begin{equation*}
    e^{i\chi}n^{2T-f}(im)^f c_{a\; \bar{a}\oplus f}(f-1)\begin{cases} W_{01}\otimes W_{a\; \bar{a}\oplus f}\\
   W_{a\; \bar{a}\oplus f} \otimes W_{10} 
    \end{cases}\,.
\end{equation*}
However, when the relative position does not change, there can be either two interactions or none. The multiplicity of the diagrams contributing to one case or the other depends on the initial and final conditions. Loosely speaking, it depends on how far apart the particles are at the beginning and at the end of the process. Thus the transition matrix associated with interacting diagrams takes the form:
\begin{equation*}
    e^{2i\chi}n^{2T-f}(im)^f \mathcal{C}(x_{in},x_{out},y_{in},y_{out}) \begin{cases} W_{01}\otimes W_{a\; \bar{a}\oplus f}\\
   W_{a\; \bar{a}\oplus f} \otimes W_{10} 
    \end{cases}\,,
\end{equation*}
with \(\mathcal{C}(x_{in},x_{out},y_{in},y_{out})\) a suitable function accounting for the multiplicity of paths contributing to the process. We computed explicitly the case \((1,2)\) (case \((1,1)\) is trivial, since \(c_{ab}(1)=1 \; \forall a,b\in\left\{0,1\right\}\)). 

\vspace{0.3cm}
\begin{center}
\begin{minipage}{\textwidth}
\begin{minipage}{0.3\textwidth} 
\includegraphics[width=0.95\textwidth]{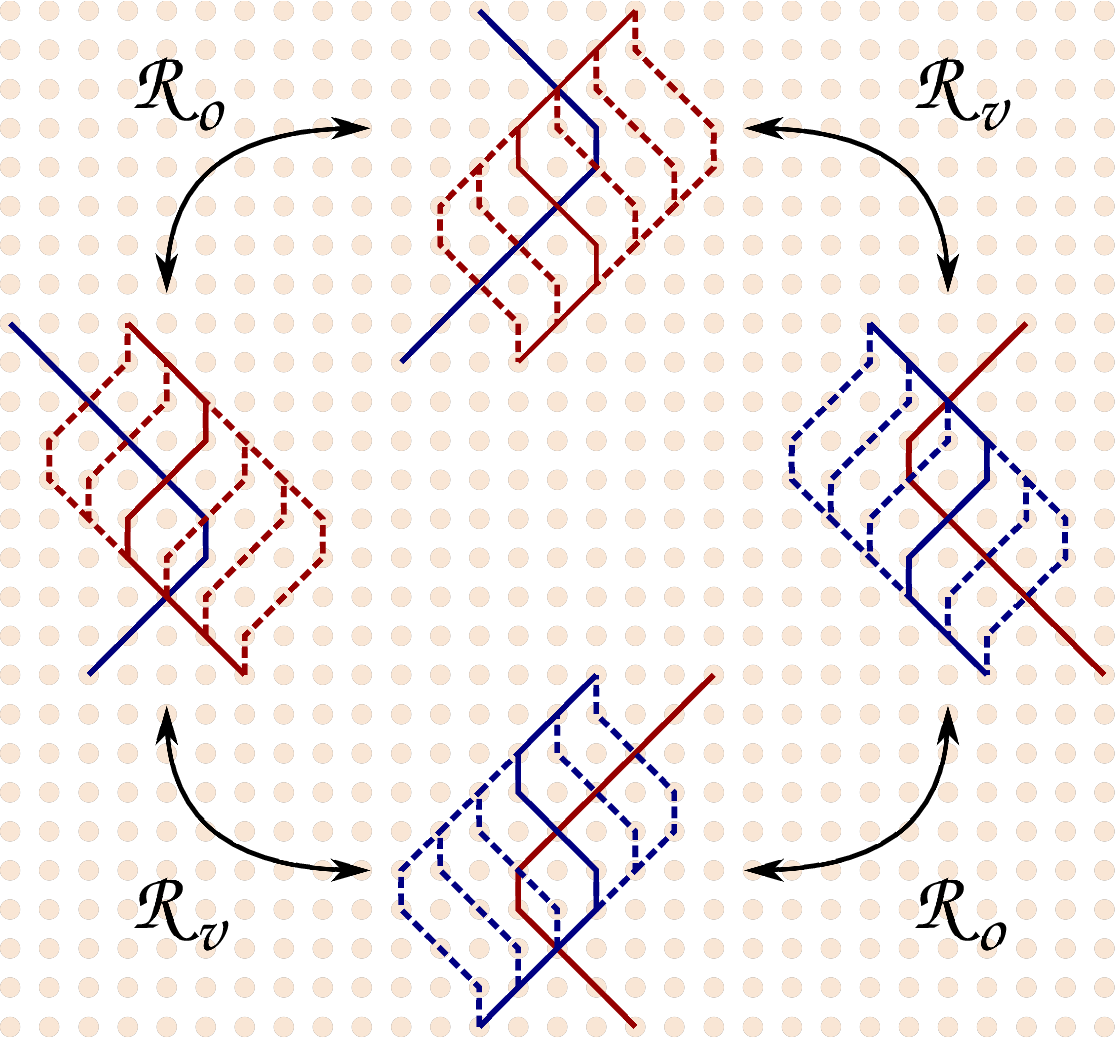} 
\end{minipage} 
\begin{minipage}{0.2\textwidth} 
\[-ie^{2i\chi}n^{2T-3}m^3\begin{cases}
    &\left[c_{11}(y_{in},y_{out},2)-\frac{|x_{in}-y_{in}|}{2}\right](W_{01}\otimes W_{11})\\
    &\left[c_{00}(y_{in},y_{out},2)-\frac{|x_{out}-y_{out}|}{2}\right](W_{01}\otimes W_{00})\\
    &\left[c_{11}(x_{in},x_{out},2)-\frac{|x_{out}-y_{out}|}{2}\right](W_{11}\otimes W_{10})\\
    &\left[c_{00}(x_{in},x_{out},2)-\frac{|x_{in}-y_{in}|}{2}\right](W_{00}\otimes W_{10})\\
\end{cases}
\]
\end{minipage}  
\end{minipage}
\end{center}
\vspace{0.3cm}

This concludes the analysis of the low-mass regime. The manual enumeration of all admissible diagrams becomes highly challenging when increasing the number of \(F\)-type transitions. Up to \(f=3\), all the diagrams we considered contained at least one path with multiplicity \(1\), namely a light-like path or a path containing a single \(F\)-type transition, thus facilitating the combinatorial analysis. However, this is no longer the case for higher perturbative orders. For instance, the class \(f=4\) contains the sub-class \((2,2)\) where neither path has multiplicity \(1\). This results in a non-trivial dependence of the coefficients on the initial and final sites, which characterises sub-classes  \((f_1,f_2)\) of all subsequent orders \(f=f_1+f_2\ge 4\) with \(f_1,f_2\ge 2\).

\subsection{High-mass limit}

We now turn to the analysis of diagrams describing the evolution of particles in the limit \(m\simeq 1\). In such a regime paths are made up almost entirely of \(F\)-type transitions, namely they trace a limited number of shifts. As a consequence, for the particles to be able to interact, both their initial and final sites have to be somewhat close. Indeed, both initial and final sites cannot be separated by a number of sites greater than the number of non-\(F\) transitions. To make this observation formal, Upon denoting with \(T\) the total time steps, let $\delta_0$ and $\delta_T$ be the relative displacement between the starting and ending points, respectively. We consider the classes of paths \(f=2T,2T-1,2T-2\), which can be further divided into sub-classes \((f_1,f_2,\delta_0=a,\delta_T=b)\) where 
\[a,b\in\mathbb{Z} \;\;\; \mbox{s.t.} \;\;\; |a|+|b|=2T-f\,.\]
To lighten the notation, when writing the transition matrices we will henceforth omit the multiplying factor containing the parameters \(m,n\), as it is the same for every diagram in the same class and can be easily recovered from \(f\) as \(n^{2T-f}(im)^f\).

Let us begin by considering the trivial case \(f=2T\), in which both particles never shift, resulting in space-time paths represented by straight vertical lines. Therefore, the particles either interact at each time step or they do not at all. Since we are only interested in interacting diagrams, the only possibility is that both paths share the same initial and final sites, resulting in the following transition matrix:
\begin{equation*}
e^{iT\chi}W_{a\;a\oplus T}\otimes W_{\bar{a}\; \bar{a}\oplus T}\,.   
\end{equation*}
Moving to the case \(f=2T-1\),  one particle is allowed one shift, while the other never shifts. Therefore,  they must either share the same initial site and end up one site apart or vice-versa. Notice that, as opposed to the low-mass regime, upon fixing the boundary conditions, different paths contain a different number of interactions

\vspace{0.3cm}
\begin{minipage}{\textwidth}
\begin{minipage}{0.5\textwidth} 
\includegraphics[width=0.55\textwidth]{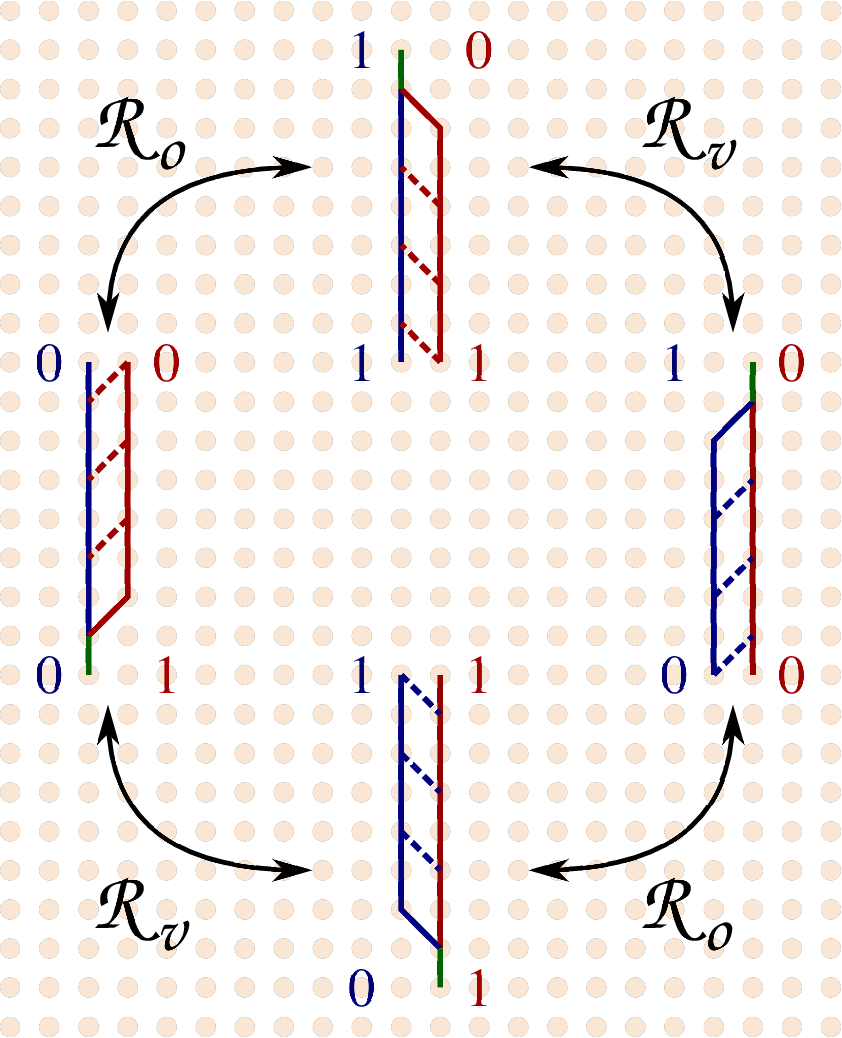} 
\end{minipage} 
\begin{minipage}{0.2\textwidth} 

\[\sum_{k=1}^\frac{T}{2}e^{2ik\chi}\begin{cases}
    &(W_{00}\otimes W_{10})\\
      &(W_{11}\otimes W_{10})\\
        &(W_{01}\otimes W_{00})\\
          &(W_{01}\otimes W_{11})\\
\end{cases}\,.\]
\end{minipage}  
\end{minipage}
\vspace{0.1cm}

Lastly, we considered the diagrams with \(f=2T-2\). The most remarkable challenge encountered in their classification resides in the non-trivial dependence of the number of interactions and the multiplicity of diagrams with a fixed number of interactions on the boundary conditions, meaning not only that different paths contain a different number of interactions (which was already the case for \(f=2T-1\)), but the multiplicity of these paths also varies from one another. This results in various sub-classes, which share no apparent common features.

\vspace{0.3cm}
\begin{minipage}{\textwidth}
\begin{minipage}{0.4\textwidth} 
\includegraphics[width=0.7\textwidth]{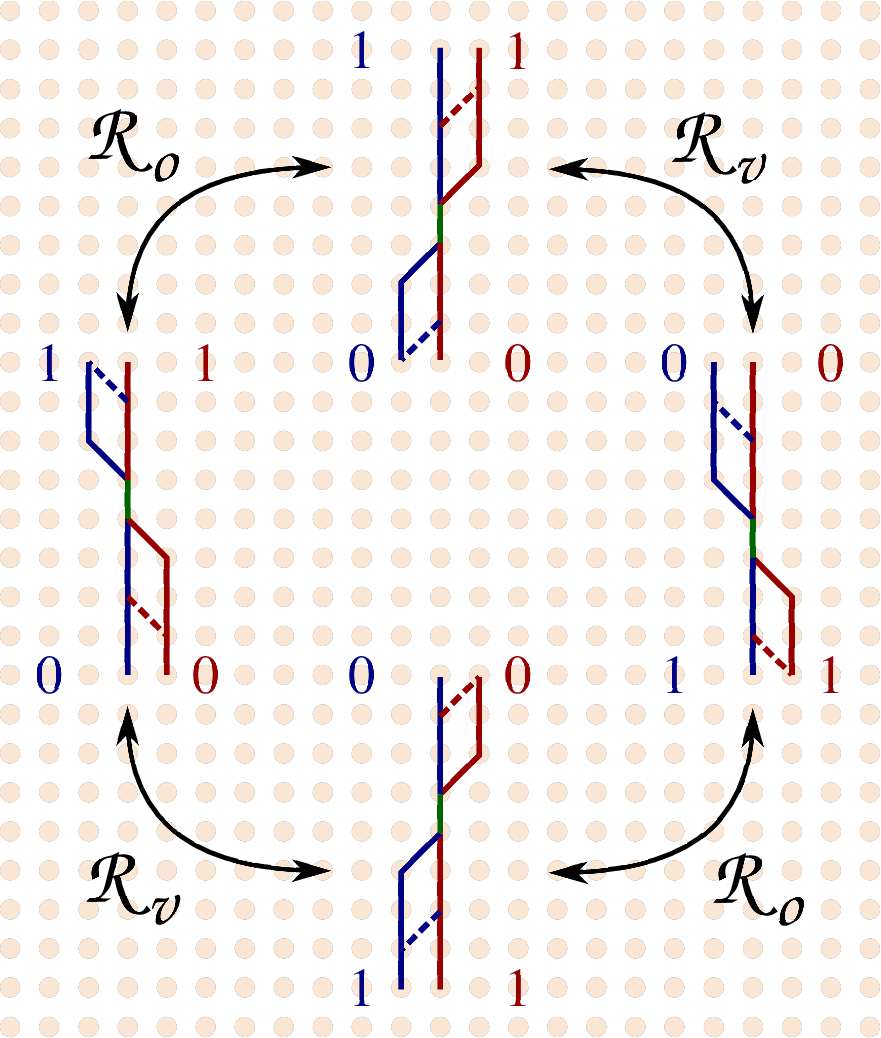} 
\end{minipage} \hspace{0.3cm}
\begin{minipage}{0.4\textwidth} 
\[\sum_{k=1}^{\frac{T}{2}-1}\left(\frac{T-2k}{2}\right)e^{2ki\chi}\begin{cases}
&(W_{01}\otimes W_{01})\\
&(W_{10}\otimes W_{10})\\
\end{cases}\]
\end{minipage}  
\end{minipage}
\vspace{0.3cm}

This concludes the analysis of the high-mass regime. From here on, the complex behaviour of the paths makes the combinatorial analysis extremely challenging, preventing the manual classification of the diagrams. 

\section{Multi-particle sector}\label{three}
\begin{figure}[h]
    \centering
    \includegraphics[width=0.15\textwidth]{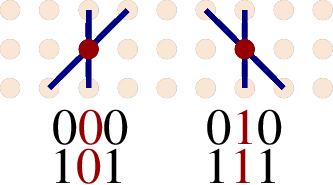}
    \caption{Example of paths violating Pauli’s exclusion principle. Highlighted in red is the site where the particles have the same internal degree of freedom.}
    \label{fig:nopauli}
\end{figure}

In this section, we briefly discuss the validity of our results when attempting to generalise them to the multi-particle sector of the Thirring QCA, focusing, in the end, on the three-particle case. The following Lemma stems directly from Lemma \ref{lemcnec}.
\begin{Lemma}\label{3part}
Within a $N$-particle process, if \(N-1\) particles can interact with the remaining one, then they cannot interact with each other.
\end{Lemma}
As a direct consequence of this, the multi-particle case presents a relevant additional difficulty, which we did not have to consider in the two-particle sector. When only two particles are evolving on the lattice, as long as they satisfy the necessary condition for an interaction to occur (Lemma \ref{lemcnec}), the results we proved in section \ref{mathtools} ensure that, whenever both particles are on the same site, they have opposite internal degrees of freedom. However, this is no longer true when dealing with more than two particles. In fact, it is possible for two particles that cannot interact with each other to arrive at the same site with the same internal degree of freedom, thus violating Pauli's exclusion principle (Fig. \ref{fig:nopauli}). 
This feature makes the perturbative approach in the number of interactions ineffective in the multi-particle sector. Indeed, we recall that the algorithmic procedure we devised stems from separating the notions of intersection between paths and interaction between particles, giving up control over the intersections of the paths in between interactions. Consequently, the algorithm in the $N \geq 3$ particle case would not be able to exclude unphysical paths violating Pauli's principle.

Contrarily, the perturbative approach in the mass parameter can still be carried out, at least in principle. Lemma \ref{3part} proves useful within our classification procedure, especially when restricted to the three-particle case. Indeed, only one of them can interact with both other particles. The classification is based on the same criteria we used for the two-particle case, but now each sub-class is identified by three integers \((f_1,f_2,f_3)\), such that \(f_1+f_2+f_3=f\), with \(f\) being the chosen perturbative order.
As for the two-particle case, this procedure is feasible as long as we restrict to the first few perturbative orders but, unfortunately, does not yield a general unique expression for the transition matrix. Moreover, the fermionic statistics of the modes imply that one should carry out the same classification procedure also for diagrams containing occurrences as in Fig. \ref{fig:nopauli}, which must then be excluded to compute the total transition matrix.

\section{Summary and conclusions}
The main objective of this work was to explore approximation techniques for the QCA approach to the theory of interacting quantum systems, focusing on the Thirring QCA. The latter is characterised by the most general on-site number-preserving interaction. Within the framework of the path sum formalism, particularly adopting a perturbative approach, we aimed to devise procedures for predicting relevant physical quantities, particularly the transition matrix for a given process. 

We started by building on the free theory, obtaining a few results valid for any on-site number-preserving interaction in one spatial dimension. We first showed that all binary strings corresponding to paths with given boundary conditions share the same weight. This implies that fixing the particle's state at a given point in space-time uniquely determines its state at any other site in the space-time lattice. Upon introducing the interaction, we identified a necessary condition for initial states to comply with Pauli's exclusion principle. Notably, particles satisfying this condition are distinguishable, as the final state of one cannot be the final state of the other. 

To address the inherent complexity of the problem, we adopted a perturbative approach in the mass parameter, which allowed us to compute the transition matrix from some basic processes. However, our results primarily yielded a classification of all allowed diagrams in the borderline regimes of extremely light or extremely heavy particles, rather than their general characterisation. Future work could extend the classification to higher perturbative orders, which might give useful insights into finding a way to reconcile the two regimes.

This work highlighted the notable challenges posed by the QCA approach to an interacting theory, even with a simple interaction model restricted to one spatial dimension. The difficulties encountered in adopting the path sum formalism suggest it might not be the most suited framework for such an endeavor. However, this may be due to our current limited understanding of the QCA approach to interaction theory. We currently have analytical solutions only for the two-particle sector of the Thirring QCA, and its integrability for an arbitrary number of particles remains an open question. Furthermore, the peculiar structure of the discrete propagator in the path sum approach to the Dirac QCA indicates strong model dependence, meaning further investigations could lead to alternative strategies that exploit features of the model we are not yet aware of. Thus, the path sum approach may still prove valuable once future research provides us with more insights, enabling us to better handle the computational challenges we now find insurmountable.


\funding{AB acknowledge financial support from European Union – Next Generation EU through the MUR
project Progetti di Ricerca d’Interesse Nazionale (PRIN)
DISTRUCT No. P2022T2JZ9.  PP acknowledge financial support from European Union – Next Generation EU through the PNNR MUR Project No. PE0000023-NQSTI}

\conflictsofinterest{The authors declare no conflicts of interest. The funders had no role in the design of the study, in the analyses, in the writing of the manuscript, or in the decision to publish the results.}

\appendixtitles{no} 
\appendixstart
\appendix
\section[\appendixname~\thesection]{Proofs of results}

\label{app}
\subsection*{Proof of Lemma \ref{lemmaperm}}
\begin{proof}
Since any permutation of the internal bits can be regarded as the composition of swaps between first-neighbour bits, it is useful to introduce the following exchange map \(\varphi\) defined as 
\begin{align*}
    \varphi \;\; : \;\; \left\{0,1\right\}^2 &\to \left\{0,1\right\}^2\\
    b_1b_2 &\mapsto \varphi(b_1b_2)=b_2b_1
\end{align*}
It is immediate to check that 
\begin{equation}
\label{Msy}
    \mathcal{M}\circ \varphi=\mathcal{M} \;\;\;\;\; \mbox{namely}\;\;\;\;\; \mathcal{M}_{b_1b_2}=\mathcal{M}_{b_2b_1} 
\end{equation} 
Thus it is sufficient to prove the statement for the case \(\Pi=\varphi\). Notice that the exchange of two bits can influence at most three elements in the corresponding string of translations, which amounts to a length-four binary sub-string, whose two central bits are the ones affected by the swap. The scenario under scrutiny can be represented as follows:
\begin{center}
\begin{tikzcd}
b_1b_2b_3b_4 \arrow{r}{\varphi} \arrow[swap]{d}{\widetilde{\mathcal{M}}} & b_1b_3b_2b_4  \arrow{d}{\widetilde{\mathcal{M}}} \\
\mathcal{M}_{b_1b_2}\mathcal{M}_{b_2b_3}\mathcal{M}_{b_3b_4} & \mathcal{M}_{b_1b_3}\mathcal{M}_{b_3b_2}\mathcal{M}_{b_2b_4}
\end{tikzcd}
\end{center}

In formula, the point of the proof is to show that
\begin{equation*}
    \Delta(\mathcal{M}_{b_1b_2}\mathcal{M}_{b_2b_3}\mathcal{M}_{b_3b_4})= \Delta(\mathcal{M}_{b_1b_3}\mathcal{M}_{b_3b_2}\mathcal{M}_{b_2b_4})
\end{equation*}
namely, by using \eqref{Msy} 
\begin{equation}
\label{Delta}
    \Delta(\mathcal{M}_{b_1 b_2})+\Delta(\mathcal{M}_{b_3b_4})=\Delta(\mathcal{M}_{b_1 b_3})+\Delta(\mathcal{M}_{b_2b_4})
\end{equation}
It is immediate to verify that for \(b_2=b_3\) Eq. \eqref{Delta} is trivially satisfied. The only relevant case is then  \(b_2=\overline{b}_3=b\), and\eqref{Delta} becomes
\begin{equation}
\label{Delta2}
    \Delta(\mathcal{M}_{b_1 b})+\Delta(\mathcal{M}_{\bar{b}b_4})=\Delta(\mathcal{M}_{b_1\bar{b}})+\Delta(\mathcal{M}_{bb_4})
\end{equation}
One then has the following possible scenarios:
\begin{itemize}
    \item \(b_1=b_4=a\): Eq. \eqref{Delta2} becomes
\begin{equation*}
    \Delta(\mathcal{M}_{a b})+\Delta(\mathcal{M}_{\bar{b}a})=\Delta(\mathcal{M}_{a\bar{b}})+\Delta(\mathcal{M}_{b a})
\end{equation*}
and it is trivially satisfied, since \(\mathcal{M}_{ab}=\mathcal{M}_{ba}\).\\
\item \(b_1=\overline{b}_4=a\): Eq. \eqref{Delta2} becomes
\begin{equation*}
    \Delta(\mathcal{M}_{a b})+\Delta(\mathcal{M}_{\bar{b}\bar{a}})=\Delta(\mathcal{M}_{a\bar{b}})+\Delta(\mathcal{M}_{b \bar{a}})
\end{equation*}
Since \(\Delta(\mathcal{M}_{\bar{a}\bar{b}})=-\Delta(\mathcal{M}_{ab})\), the equation above reduces to the identity \(0=0\).\\

\end{itemize}
Therefore, we have shown that the exchange of two neighbouring internal bits in the binary string of the internal degrees of freedom leaves the relative displacement associated with the string of translations unchanged. Consequently, as we observed at the beginning of the proof, the result holds for any permutation of the internal bits.
\end{proof}
\subsection*{Proof of Lemma \ref{lemmaweight}}

\begin{proof}
Consider the string of transitions \(s\) associated with a path connecting the site \(x_{in}\) to the site \(x_{out}\) in \(T\) time steps. Suppose its boundary conditions are given, namely the state of site \(x_{in}\) at time step \(0\) and the state of site \(x_{out}\) at time \(T\), denoted with \(b_0\) and \(b_T\), respectively. Let \(f,r\) and \(l\) denote the number of \(F,R\) and \(L\)-type transitions contained within such a string. 
To compute the weight of the corresponding binary string \(b\) associated with the internal degrees of freedom, we need to consider the following:
\begin{itemize}
    \item each \(F\) transition contributes one, each \(L\) transition contributes two, while \(R\) transitions do not contribute at all (see def. \ref{defM}).
    \item within the string of transitions \(S\), each transition identifies a pair of bits in the binary string \(b\). The first bit is shared with the previous transition and second one is shared with with the following one. Therefore, to avoid over-counting, the contribution of each transition to the weight must be divided by \(2\).  
    \item the previous observation does not hold for the first and last elements of the string of transitions, therefore the contribution of the first and last bits (which are fixed by the boundary conditions) must be accounted for before dividing
\end{itemize}
It follows that the weight of the binary string \(b\) is given by
\begin{equation}
\label{wlr}
    w(b)=\frac{2l+f+b_0+b_T}{2}
\end{equation}
By exploiting the following equations \cite{dariano2014pathint}
\begin{equation*}
    x_{in}-x_{out}=l-r \;\;\;\;\;\; T=r+l+f
\end{equation*}
eq. \eqref{wlr} becomes
\begin{equation*}
      w(b)=\frac{T+x_{in}-x_{out}+b_0+b_T}{2}
\end{equation*}
thus concluding the proof.
\end{proof}
\subsection*{Proof of Corollary \ref{coruni}}
\begin{proof}

It follows directly from eq. \eqref{pesiW}
\begin{equation*}
    w(b)=\frac{T+x_{in}-x_{out}+b_0+b_T}{2}
\end{equation*}
fixing \(x_{in}\) and \(b_0\) while treating \(x_{out}\) and \(b_T\) as variables, since the numerator must be even, one can then write
\begin{equation*}
    b_T=[T+x_{in}-x_{out}+b_0]\mbox{mod}2 \;\;\;\;\; \forall \; x_{out}
\end{equation*}
thus proving the statement.
\end{proof}
\subsection*{Proof of Lemma \ref{lemcnec}}

\begin{proof}
Let \(x_{in}\) and \(y_{in}=x_{in}+\delta_0\) represent the initial space coordinates of the two particles, denoted as \(A\) and \(B\), and let  \((v,t)\) be the space-time coordinates of the interaction site. The weights of the binary strings \(b^A\) and \(b^B\), corresponding to paths connecting \(x_{in}\) and \(y_{in}\), respectively, to \(v\) in \(t\) discrete-time steps, are given by eq. \eqref{pesiW}
\begin{align*}
    w(b_A)&=\frac{t+x_{in}-v-b_0-\alpha}{2}\\
    w(b_B)&=\frac{t+y_{in}-v-d(y_{in})-\bar{\alpha}}{2}
\end{align*}
where \(\alpha=b^A_t,\bar{\alpha}=b^B_t\) denote the (opposite) internal degrees of freedom of the particles when both are at the interaction site, according to Pauli's exclusion principle. It follows that
\begin{align*}
    b^A_0&=[t+x_{in}-v+\alpha]\mbox{mod}2\\
  b^B_0&=[t+y_{in}-v+\bar{\alpha}]\mbox{mod}2\\
    &=[t+(x_{in}+\delta_0)-v+(1+\alpha)]\mbox{mod}2\\
    &=\delta_0\mbox{mod}2\oplus[1+(t+x_{in}-v+\alpha)]\mbox{mod}2\\
    &=\delta_0\mbox{mod}2\oplus\overline{b^A_0}
\end{align*}
This leads to:
\begin{equation*}
\begin{cases}
   b^B_0=\overline{b^A_0} \;\;\;\;\;&\mbox{if}\;\;\;\;\; \delta_0 \;\;\mbox{even}\\
   b^B_0=b^A_0 \;\;\;\;\;&\mbox{if}\;\;\;\;\; \delta_0 \;\;\mbox{odd}\\
\end{cases}
\end{equation*}
concluding the proof.
\end{proof}

\subsection*{Proof of Corollary \ref{cor}}
\begin{proof}
We can use the same argument as in the proof of Lemma \ref{lemcnec} to prove that the conditions we obtained for the initial states also hold for the final states. Therefore, upon adopting the same notation as in the previous proof and denoting  the final sites with \((x_{out},T)\) and \((y_{out},T)\), we conclude from the previous Lemma that, for \(\delta_T=x_{out}-y_{out}=0\), one must have \(b^A_T=\overline{b^B_T}\). Namely, if one particle is in state \(X\) at a given site, the other must be in state \(\overline{X}\) whenever it lands on that same site. One can then exploit such difference to distinguish the two particles.
\end{proof}
\subsection*{Proof of Lemma \ref{lemmadist}}
\begin{proof}
The proof is structured in two steps:
\begin{itemize}
    \item showing that the particle which, before the interaction, is situated at a smaller (larger) spatial coordinate arrives at the interaction site in the state \(0\) (\(1\));
    \item proving that the particle which, at the interaction site, is in the state \(0\) (\(1\)), in the next step has a spatial coordinate larger (smaller) than the other.
\end{itemize}

To prove the first statement, one can evaluate the relative displacement of each particle associated with the transition leading up to the interaction. Since, at the interaction site, the particles must have opposite internal degrees of freedom, one has 
\begin{align*}
    \Delta(\mathcal{M}_{a0})&=1-a=\bar{a}\\
    \Delta(\mathcal{M}_{b1})&=(-1)^bb
\end{align*}
The goal is to prove that, for a single interaction, the particle ending up in the state \(0\) (\(1\)) arrives from the right (left), namely that
\begin{equation}
    \label{lemmastepuno}
    \Delta(\mathcal{M}_{a0})\ge\Delta(\mathcal{M}_{b1}) \;\;\;\;\; \forall a,b\in\left\{0,1\right\}
\end{equation}
By direct inspection of the possible cases, one has
\begin{enumerate}
    \item \((a,b)=(0,0)\): \(\Delta(\mathcal{M}_{00})=1>\Delta(\mathcal{M}_{01})=0\)
    \item \((a,b)=(0,1)\): \(\Delta(\mathcal{M}_{00})=1>\Delta(\mathcal{M}_{11})=-1\)
    \item \((a,b)=(1,0)\): \(\Delta(\mathcal{M}_{10})=0=\Delta(\mathcal{M}_{01})\)
    \item \((a,b)=(1,1)\): \(\Delta(\mathcal{M}_{10})=0>\Delta(\mathcal{M}_{11})=-1\)
\end{enumerate}
Denoting with \(A\) (\(B\)) the particle ending up in the state \(0\) (\(1\)), this amounts to writing \(\delta_{t-1}\le0\), which concludes the first step.\\

To conclude the proof, it suffices to show that the particle which is in state \(0\) at the interaction site moves on to a site with a greater spatial coordinate than the other, which was in state \(1\) when interacting. This requires proving that
\[\Delta(\mathcal{M}_{0a})\ge\Delta(\mathcal{M}_{1b}) \;\;\;\;\; \forall (a,b)\in\left\{0,1\right\}\]
It follows immediately from eq. \eqref{lemmastepuno}, since \(\mathcal{M}_{xy}=\mathcal{M}_{yx}\), and implies \(\delta_{t+1}\ge0\), concluding the proof.
\end{proof}

\subsection*{Proof of Lemma \ref{3part}}
\begin{proof}
Let us denote with \(x_A,x_B,x_C\) and \(b^A_0,b^B_0,b^C_0\) the initial space coordinate and internal degree of freedom of particle \(A,B,C\), respectively. Since by hypothesis particle \(B\) interacts with both particle \(A\) and particle \(C\), we know from Lemma \ref{lemcnec} that the following conditions must hold:
\begin{equation*}
\begin{cases}
    b^B_0=\overline{b^i_0} \;\;\;\;\;&\mbox{if}\;\;\;\;\; \delta^i_0 \;\;\mbox{even}\\
    b^B_0=b_0^i \;\;\;\;\;&\mbox{if}\;\;\;\;\; \delta_0^i \;\;\mbox{odd}\\
\end{cases} \;\;\;\;\; i=A,C
\end{equation*}
where \(\delta_0^A\) and \(\delta_0^C\) denote the relative position of particles \(A\) and \(C\) with respect to particle \(B\), namely \(\delta_0^i\coloneqq x_B-x_i\), \(i=A,C\). The relative position between particles \(A\) and \(C\) can thus be evaluated as \(\Delta\coloneqq x_A-x_C=\delta_0^C-\delta_0^A\) (the sign is irrelevant since we are only interested in its parity). All possible scenarios are summarised in table \ref{table1}.

\begin{table}[h]
    \centering
    \begin{tabular}{|c|c|c|} \hline  
         &  \(\delta_0^C\) even&  \(\delta_0^C\) odd\\ \hline  
         \(\delta_0^A\) even&  \(\Delta\) even \(\wedge\) \(\overline{b_0^A}=\overline{b_0^C}\)&  \(\Delta\) odd \(\wedge\) \(\overline{b_0^A}=b_0^C\)\\ \hline  
         \(\delta_0^A\) odd&  \(\Delta\) odd \(\wedge\) \(b_0^A=\overline{b_0^C}\)&  \(\Delta\) even \(\wedge\) \(b_0^A=b_0^C\)\\ \hline 
    \end{tabular}
 \caption{Relations between the relative positions of particles \(A\) and \(C\) with respect to B and their internal degrees of freedom.   \label{table1}}
\end{table}

It is immediate to check that no entry of the table satisfies the necessary condition for particles \(A\) and \(C\) to be able to interact, thus concluding the proof.
\end{proof}

\begin{adjustwidth}{-\extralength}{0cm}

\reftitle{References}


\bibliography{Thesis_bibliography}

\begin{thebibliography}{999}

\bibitem[Feynman(1981)]{feynman1981simulating}
Feynman, R.P.
\newblock Simulating physics with computers, 1981.
\newblock {\em International Journal of Theoretical Physics} {\bf 1981}, {\em
  21}.

\bibitem[Deutsch(1985)]{deutsch1985quantum}
Deutsch, D.
\newblock Quantum theory, the Church--Turing principle and the universal
  quantum computer.
\newblock {\em Proceedings of the Royal Society of London. A. Mathematical and
  Physical Sciences} {\bf 1985}, {\em 400},~97--117.

\bibitem[Gr\"ossing and Zeilinger(1988)]{qcagrozei}
Gr\"ossing, G.; Zeilinger, A.
\newblock Quantum Cellular Automata.
\newblock {\em Complex Systems} {\bf 1988}, {\em 2},~197--208.

\bibitem[Farrelly(2020)]{Farrelly_2020}
Farrelly, T.
\newblock A review of Quantum Cellular Automata.
\newblock {\em Quantum} {\bf 2020}, {\em 4},~368.
\newblock {\url{https://doi.org/10.22331/q-2020-11-30-368}}.

\bibitem[Preskill(2018)]{preskill2018quantum}
Preskill, J.
\newblock Quantum computing in the NISQ era and beyond.
\newblock {\em Quantum} {\bf 2018}, {\em 2},~79.

\bibitem[Arrighi(2019)]{arrighi2019overview}
Arrighi, P.
\newblock An overview of quantum cellular automata.
\newblock {\em Natural Computing} {\bf 2019}, {\em 18},~885--899.

\bibitem[Bialynicki-Birula(1994)]{PhysRevD.49.6920}
Bialynicki-Birula, I.
\newblock Weyl, Dirac, and Maxwell equations on a lattice as unitary cellular
  automata.
\newblock {\em Phys. Rev. D} {\bf 1994}, {\em 49},~6920--6927.
\newblock {\url{https://doi.org/10.1103/PhysRevD.49.6920}}.

\bibitem[B{\'e}ny(2017)]{beny2017inferring}
B{\'e}ny, C.
\newblock Inferring effective field observables from a discrete model.
\newblock {\em New Journal of Physics} {\bf 2017}, {\em 19},~013013.

\bibitem[Osborne(2019)]{osborne2019continuum}
Osborne, T.J.
\newblock Continuum limits of quantum lattice systems.
\newblock {\em arXiv preprint arXiv:1901.06124} {\bf 2019}.

\bibitem[Perinotti(2020)]{Perinotti_2020}
Perinotti, P.
\newblock Quantum Field Theory from first principles.
\newblock {\em Istituto Lombardo - Accademia di Scienze e Lettere - Rendiconti
  di Scienze} {\bf 2020}.
\newblock {\url{https://doi.org/10.4081/scie.2017.649}}.

\bibitem[Mlodinow and Brun(2021)]{mlodinow2021fermionic}
Mlodinow, L.; Brun, T.A.
\newblock Fermionic and bosonic quantum field theories from quantum cellular
  automata in three spatial dimensions.
\newblock {\em Physical Review A} {\bf 2021}, {\em 103},~052203.

\bibitem[Zimbor{\'a}s et~al.(2022)Zimbor{\'a}s, Farrelly, Farkas, and
  Masanes]{zimboras2022does}
Zimbor{\'a}s, Z.; Farrelly, T.; Farkas, S.; Masanes, L.
\newblock Does causal dynamics imply local interactions?
\newblock {\em Quantum} {\bf 2022}, {\em 6},~748.

\bibitem[Eon et~al.(2023)Eon, Di~Molfetta, Magnifico, and
  Arrighi]{eon2023relativistic}
Eon, N.; Di~Molfetta, G.; Magnifico, G.; Arrighi, P.
\newblock A relativistic discrete spacetime formulation of 3+ 1 QED.
\newblock {\em Quantum} {\bf 2023}, {\em 7},~1179.

\bibitem[D'Ariano and Perinotti(2014)]{PhysRevA.90.062106}
D'Ariano, G.M.; Perinotti, P.
\newblock Derivation of the Dirac equation from principles of information
  processing.
\newblock {\em Phys. Rev. A} {\bf 2014}, {\em 90},~062106.
\newblock {\url{https://doi.org/10.1103/PhysRevA.90.062106}}.

\bibitem[Georgescu et~al.(2014)Georgescu, Ashhab, and
  Nori]{georgescu2014quantum}
Georgescu, I.M.; Ashhab, S.; Nori, F.
\newblock Quantum simulation.
\newblock {\em Reviews of Modern Physics} {\bf 2014}, {\em 86},~153--185.

\bibitem[Arrighi et~al.(2020)Arrighi, B{\'e}ny, and Farrelly]{Arrighi_2020}
Arrighi, P.; B{\'e}ny, C.; Farrelly, T.
\newblock A quantum cellular automaton for one-dimensional QED.
\newblock {\em Quantum Information Processing} {\bf 2020}, {\em 19}.
\newblock {\url{https://doi.org/10.1007/s11128-019-2555-4}}.

\bibitem[Farrelly and Streich(2020)]{farrelly2020discretizing}
Farrelly, T.; Streich, J.
\newblock Discretizing quantum field theories for quantum simulation.
\newblock {\em arXiv preprint arXiv:2002.02643} {\bf 2020}.

\bibitem[Bisio et~al.(2021)Bisio, Mosco, and Perinotti]{bisio2021scattering}
Bisio, A.; Mosco, N.; Perinotti, P.
\newblock Scattering and perturbation theory for discrete-time dynamics.
\newblock {\em Physical Review Letters} {\bf 2021}, {\em 126},~250503.

\bibitem[Wilson(1983)]{wilson1983renormalization}
Wilson, K.G.
\newblock The renormalization group and critical phenomena.
\newblock {\em Reviews of Modern Physics} {\bf 1983}, {\em 55},~583.

\bibitem[Arrighi et~al.(2018)Arrighi, Di~Molfetta, and Eon]{Arrighi_2018}
Arrighi, P.; Di~Molfetta, G.; Eon, N., A Gauge-Invariant Reversible Cellular
  Automaton.
\newblock In {\em Lecture Notes in Computer Science}; Springer International
  Publishing,  2018; pp. 1--12.
\newblock {\url{https://doi.org/10.1007/978-3-319-92675-9_1}}.

\bibitem[Arnault et~al.(2016)Arnault, Di~Molfetta, Brachet, and
  Debbasch]{arnault2016quantum}
Arnault, P.; Di~Molfetta, G.; Brachet, M.; Debbasch, F.
\newblock Quantum walks and non-Abelian discrete gauge theory.
\newblock {\em Physical Review A} {\bf 2016}, {\em 94},~012335.

\bibitem[Centofanti et~al.(2024)Centofanti, Bisio, and
  Perinotti]{centofanti2024}
Centofanti, E.; Bisio, A.; Perinotti, P.
\newblock A massless interacting Fermionic Cellular Automaton exhibiting bound
  states,  2024,  \href{http://arxiv.org/abs/2304.14687}{{\normalfont
  [arXiv:quant-ph/2304.14687]}}.

\bibitem[Bibeau-Delisle et~al.(2015)Bibeau-Delisle, Bisio, D'Ariano, Perinotti,
  and Tosini]{bibeau2015doubly}
Bibeau-Delisle, A.; Bisio, A.; D'Ariano, G.M.; Perinotti, P.; Tosini, A.
\newblock Doubly special relativity from quantum cellular automata.
\newblock {\em Europhysics Letters} {\bf 2015}, {\em 109},~50003.

\bibitem[Bisio et~al.(2016)Bisio, D'Ariano, and Perinotti]{Bisio_2016}
Bisio, A.; D'Ariano, G.M.; Perinotti, P.
\newblock Special relativity in a discrete quantum universe.
\newblock {\em Physical Review A} {\bf 2016}, {\em 94}.
\newblock {\url{https://doi.org/10.1103/physreva.94.042120}}.

\bibitem[Arrighi and Martiel(2017)]{arrighi2017quantum}
Arrighi, P.; Martiel, S.
\newblock Quantum causal graph dynamics.
\newblock {\em Physical Review D} {\bf 2017}, {\em 96},~024026.

\bibitem[Apadula et~al.(2020)Apadula, Bisio, D'ariano, and
  Perinotti]{apadula2020symmetries}
Apadula, L.; Bisio, A.; D'ariano, G.M.; Perinotti, P.
\newblock Symmetries of the Dirac quantum walk and emergence of the de Sitter
  group.
\newblock {\em Journal of Mathematical Physics} {\bf 2020}, {\em 61}.

\bibitem[Schild(1948)]{PhysRev.73.414}
Schild, A.
\newblock Discrete Space-Time and Integral Lorentz Transformations.
\newblock {\em Phys. Rev.} {\bf 1948}, {\em 73},~414--415.
\newblock {\url{https://doi.org/10.1103/PhysRev.73.414}}.

\bibitem[Debbasch(2019)]{debbasch2019action}
Debbasch, F.
\newblock Action principles for quantum automata and Lorentz invariance of
  discrete time quantum walks.
\newblock {\em Annals of Physics} {\bf 2019}, {\em 405},~340--364.

\bibitem[Arrighi et~al.(2014)Arrighi, Facchini, and
  Forets]{arrighi2014discrete}
Arrighi, P.; Facchini, S.; Forets, M.
\newblock Discrete Lorentz covariance for quantum walks and quantum cellular
  automata.
\newblock {\em New Journal of Physics} {\bf 2014}, {\em 16},~093007.

\bibitem[D'Ariano et~al.(2014)D'Ariano, Mosco, Perinotti, and
  Tosini]{dariano2014pathint}
D'Ariano, G.M.; Mosco, N.; Perinotti, P.; Tosini, A.
\newblock Path-integral solution of the one-dimensional Dirac quantum cellular
  automaton.
\newblock {\em Physics Letter A} {\bf 2014}, {\em 378},~3165--3168.

\bibitem[Bisio et~al.(2018)Bisio, D'Ariano, Perinotti, and
  Tosini]{bisio2018thirring}
Bisio, A.; D'Ariano, G.M.; Perinotti, P.; Tosini, A.
\newblock Thirring quantum cellular automaton.
\newblock {\em Physical Review A} {\bf 2018}, {\em 97},~032132.

\bibitem[Gross et~al.(2012)Gross, Nesme, Vogts, and Werner]{gross2012index}
Gross, D.; Nesme, V.; Vogts, H.; Werner, R.F.
\newblock Index theory of one dimensional quantum walks and cellular automata.
\newblock {\em Communications in Mathematical Physics} {\bf 2012}, {\em
  310},~419--454.

\bibitem[Fidkowski et~al.(2019)Fidkowski, Po, Potter, and
  Vishwanath]{PhysRevB.99.085115}
Fidkowski, L.; Po, H.C.; Potter, A.C.; Vishwanath, A.
\newblock Interacting invariants for Floquet phases of fermions in two
  dimensions.
\newblock {\em Phys. Rev. B} {\bf 2019}, {\em 99},~085115.
\newblock {\url{https://doi.org/10.1103/PhysRevB.99.085115}}.

\bibitem[Freedman and Hastings(2020)]{freedman2020classification}
Freedman, M.; Hastings, M.B.
\newblock Classification of quantum cellular automata.
\newblock {\em Communications in Mathematical Physics} {\bf 2020}, {\em
  376},~1171--1222.

\bibitem[Haah et~al.(2022)Haah, Fidkowski, and Hastings]{haah2022nontrivial}
Haah, J.; Fidkowski, L.; Hastings, M.B.
\newblock Nontrivial quantum cellular automata in higher dimensions.
\newblock {\em Communications in Mathematical Physics} {\bf 2022}, pp. 1--72.

\bibitem[Bisio et~al.(2015)Bisio, D'Ariano, and Tosini]{BISIO2015244}
Bisio, A.; D'Ariano, G.M.; Tosini, A.
\newblock Quantum field as a quantum cellular automaton: The Dirac free
  evolution in one dimension.
\newblock {\em Annals of Physics} {\bf 2015}, {\em 354},~244--264.
\newblock {\url{https://doi.org/https://doi.org/10.1016/j.aop.2014.12.016}}.

\bibitem[Cayley(1878)]{cayley}
Cayley, P.
\newblock Desiderata and Suggestions: No. 2. The Theory of Groups: Graphical
  Representation.
\newblock {\em American Journal of Mathematics} {\bf 1878}, {\em 1},~174--176.

\bibitem[de~La~Harpe(2000)]{de2000topics}
de~La~Harpe, P.
\newblock {\em Topics in geometric group theory}; University of Chicago Press,
  2000.

\bibitem[Lieb and Wu(1968)]{PhysRevLett.20.1445}
Lieb, E.H.; Wu, F.Y.
\newblock Absence of Mott Transition in an Exact Solution of the Short-Range,
  One-Band Model in One Dimension.
\newblock {\em Phys. Rev. Lett.} {\bf 1968}, {\em 20},~1445--1448.
\newblock {\url{https://doi.org/10.1103/PhysRevLett.20.1445}}.

\bibitem[Coleman(1975)]{PhysRevD.11.2088}
Coleman, S.
\newblock Quantum sine-Gordon equation as the massive Thirring model.
\newblock {\em Phys. Rev. D} {\bf 1975}, {\em 11},~2088--2097.
\newblock {\url{https://doi.org/10.1103/PhysRevD.11.2088}}.

\bibitem[Korepin(1979)]{Korepin:1979qq}
Korepin, V.E.
\newblock {Direct Calculation of the S matrix in the massive Thirring model}.
\newblock {\em Theor. Math. Phys.} {\bf 1979}, {\em 41},~953--967.
\newblock {\url{https://doi.org/10.1007/BF01028501}}.

\bibitem[Essler et~al.(2005)Essler, Frahm, G{\"o}hmann, Kl{\"u}mper, and
  Korepin]{essler2005one}
Essler, F.H.; Frahm, H.; G{\"o}hmann, F.; Kl{\"u}mper, A.; Korepin, V.E.
\newblock {\em The one-dimensional Hubbard model}; Cambridge University Press,
  2005.

\bibitem[Hubbard and Flowers(1963)]{doi:10.1098/rspa.1963.0204}
Hubbard, J.; Flowers, B.H.
\newblock Electron correlations in narrow energy bands.
\newblock {\em Proceedings of the Royal Society of London. Series A.
  Mathematical and Physical Sciences} {\bf 1963}, {\em 276},~238--257,
  \href{http://arxiv.org/abs/https://royalsocietypublishing.org/doi/pdf/10.1098/rspa.1963.0204}{{\normalfont
  [https://royalsocietypublishing.org/doi/pdf/10.1098/rspa.1963.0204]}}.
\newblock {\url{https://doi.org/10.1098/rspa.1963.0204}}.

\bibitem[Thirring(1958)]{THIRRING195891}
Thirring, W.E.
\newblock A soluble relativistic field theory.
\newblock {\em Annals of Physics} {\bf 1958}, {\em 3},~91--112.
\newblock {\url{https://doi.org/https://doi.org/10.1016/0003-4916(58)90015-0}}.

\bibitem[\"Ostlund and Mele(1991)]{PhysRevB.44.12413}
\"Ostlund, S.; Mele, E.
\newblock Local canonical transformations of fermions.
\newblock {\em Phys. Rev. B} {\bf 1991}, {\em 44},~12413--12416.
\newblock {\url{https://doi.org/10.1103/PhysRevB.44.12413}}.

\end{thebibliography}

%


\PublishersNote{}
\end{adjustwidth}
\end{document}